%% file: dstc_arxiv.tex
\documentclass[journal]{IEEEtran}

\usepackage{epsfig}
\usepackage{url}
\usepackage{latexsym}
\usepackage{amssymb}
\usepackage{amsbsy} 
\usepackage{amsmath} 
\usepackage{float}
\usepackage{makeidx}
\usepackage{graphicx}
\usepackage{syntonly}

\begin{document}


\input{journal_paper_prelude}           
\input{sec_intro}            
\input{sec_sysmod}           
\input{sec_protocols}        
\input{sec_simulations}       
\input{sec_conclusion}       


\bibliography{dstc}              
\bibliographystyle{ieeetr} 
\end{document}

%% file: journal_paper_prelude.tex

\title{DSTC Layering Protocols in Wireless Relay Networks}
\author{Pannir~Selvam~Elamvazhuthi,~Parag~Shankar~Kulkarni, and~Bikash~Kumar~Dey,~\IEEEmembership{Member,~IEEE}
\thanks{This work has been carried out at Indian Institute of Technology Bombay (IITB), Mumbai,India.} 
\thanks{P.S. Elamvazhuthi, a research scholar at IITB, India, is with Cognizant Technology Solutions India Pvt. Ltd., Chennai, India (e-mail: epannirselvam@ee.iitb.ac.in).} 
\thanks{P.S. Kulkarni, after completing M.Tech. at IITB in 2008, working for Juniper Networks Inc., Bangalore, India (e-mail: miparag@gmail.com).} 
\thanks{B.K. Dey is with the Department of Electrical Engineering, IITB (e-mail: bikash@ee.iitb.ac.in).}}
 
\maketitle

%
\begin{abstract}
\input{abstract}
\end{abstract}

%% file: abstract.tex
With multiple antennas at transmitter and receiver, rate of transmission and reliability of information are improved. When there is no possibility of increasing the number of antennas, for example in mobile handsets, sensor networks, etc., the benefits of multiple antenna systems are obtained by cooperation amongst individual \emph{radio nodes}. \\
\indent In literature, cooperation amongst two users having single antenna each, attempting to send independent data to the same destination has been studied by many authors and various strategies have been formulated.  Studies have been carried out to use relays with single antenna each, to convey information from a single source to a destination.  Distributed space-time coding has been proposed which does not require orthogonal channels to be allocated to various transmitting units, leading to better utilization of the spectrum.  Some latest literature analyze cases when the relays have multiple antennas also.\\
\indent Our system model consists of a source-destination pair with two layers of relays in which `weaker' links between source and second layer and between the first layer and destination are also considered.  We propose five different protocols out of which one is a straight forward extension of an existing system, which is used for comparison.\\
\indent We have derived the signal-to-noise ratio at the destination for all the protocols and by maximizing this, found the optimum power to be allocated to various relay and source transmissions.  We also show that under reasonable channel strength of the `weaker' links, the proposed protocols perform ($ \approx 2$ dB) better than the existing basic protocol.  As expected, the degree of improvement increases with the strength of the weaker links.
\indent We have also shown that if receive channel knowledge is available with 50\% of the relays, reliability and data rate can be increased by adopting a technique proposed in this paper.

%% file: sec_intro.tex
\section{Introduction}
The enormous potential of multiple-input multiple-output (MIMO) communications has drawn considerable attention from the scientific community since more than a decade.  Two of the benefits of MIMO are the spatial multiplexing and diversity gains over that of single antenna systems. Installing more antennas in a small equipment may be infeasible due to space constraints.  Therefore virtual MIMO systems constituting multiple wireless systems started to take shape.  As the name suggests virtual MIMO is to simulate multiple-antenna system without having one. This can be achieved by cooperation amongst multiple radio nodes and is known as \emph{cooperative communication}, as the radio nodes cooperate with each other to obtain virtual MIMO.  \\
\indent In literature, cooperation amongst two users having single antenna each, attempting to send independent data to the same destination has been studied by many authors and various strategies have been formulated.  Studies have been carried out to use relays with single antenna each, to convey information from a single source to one destination.  Distributed space-time coding (DSTC) has been proposed which does not require orthogonal channels to be allocated to various transmitting units with single antenna each, leading to better utilization of the spectrum.\\
\indent The choice of the right space-time coding (STC) in a distributed fashion depends on the requirement.  Orthogonal space-time block coding (OSTBC) \cite{tarokh} can be selected to maximize the diversity gain and minimize the receiver complexity.  Codes \cite{gamal,sethuraman,heath} that maximize both diversity gain and transmission rate, but with a rather high receiver complexity are also available to choose from.  Bell Laboratories layered space-time (BLAST) codes \cite{wolniansky} or codes with trace-orthogonal design (TOD) \cite{barbarossa} can also be selected.  BLAST codes  maximize the rate, sacrificing part of the diversity gain, but with intermediate receiver complexity and the TOD has a flexible way to trade complexity, bit rate, and bit error rate.\\
\indent Sendonaris et al. considered a system (\cite{sendonaris}) with one destination and two sources cooperating with each other to achieve better performance.  DSTC proposed by Laneman and Wornell, used a \emph{space-time} code (\cite{laneman}) at relays and achieved higher spectral efficiency than repetition-based schemes.  Jing and Hassibi used a system (\cite{jing}) with a layer of relays between source and destination and obtained the benefits of DSTC.  Borade et al. used multiple layers of radio nodes (\cite{borade}) to relay information from source to destination.  Here the weaker links between non-consecutive layers of nodes were neglected.
Amplify-and-forward (e.g. \cite{borade}), decode-and-forward (e.g. \cite{laneman}), coded cooperation (e.g. \cite{hunter}), and simple process-and-forward (e.g. \cite{jing}) are some of the strategies used.\\
\indent In this paper, we consider a multihop network, as shown in Fig.  \ref{fig:sysmodel}, of \emph{single-antenna} radio nodes with two layers of relays between source and destination.  We adopt the strategy of simple processing and forwarding at the relays proposed by Jing and Hassibi in \cite{jing}.  However we also make use of the weaker links between the non-consecutive layers shown by dashed lines in Fig. \ref{fig:sysmodel}.
\subsection{Motivation} \label{motivation}
\indent In the previous works, the channels from source or one layer of relays to the next layer of relays or destination were considered to have same power loss, whereas the channel from a radio node to any other radio node (not in the next layer) was considered to have zero gain.  We assume that these channels (we shall call them `weak') have a smaller but non zero gain.\\
\indent  We consider schemes which make use of these `weak' signals as well.  After comparing these schemes using simulations, we come up with simple guidelines to select an appropriate scheme depending on the channel strength (power loss) and transmitted power.  We show that the proposed schemes perform better than the simple extension of the basic protocol proposed by Jing and Hassibi in \cite{jing}.  
\subsection{Contribution} \label{contribution}
We have
\begin{itemize}
\item {Proposed five different protocols for our system model and obtained maximum likelihood (ML) decoders.}
\item {Calculated the signal-to-noise ratios (SNRs) at the destination and obtained optimum power allocation for transmitters by maximizing SNR.}
\item {Analyzed and compared the performances of the proposed protocols using simulations, and shown that under reasonable strength of the `weak' channels the proposed protocols perform better than the basic protocol (\cite{jing}).}
\item {Showed, using simulations, that we can use random real orthogonal matrices instead of random complex unitary matrices employed in \cite{jing} at the relays.}
\end{itemize}
\indent This paper is organized as follows.  The system model and the previous work are detailed in Section \ref{systemmodel}.  Thereafter in Section \ref{protocolsfromJHS} the protocols derived from the basic one proposed in \cite{jing} have been analyzed, ML decoding rules have been worked out, and receive SNRs have been derived.  In Section \ref{simulations} optimum power allocations are obtained and BER plots of the protocols are compared using simulations.  Finally in Section \ref{conclusion} we draw conclusion.

%% file: sec_sysmod.tex
\section{System Model} \label{systemmodel}
\begin{figure}[!htbp]
\begin{centering}
\includegraphics[width=0.35 \textwidth]{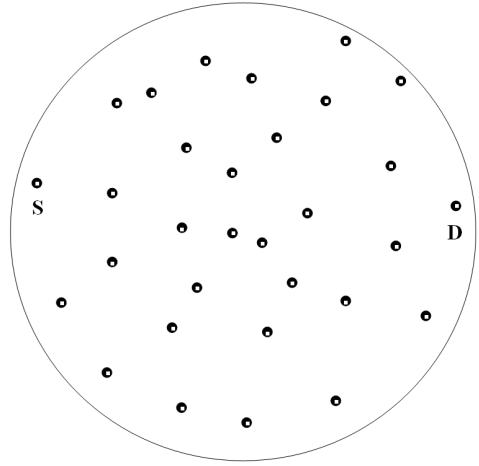}
\par
\caption{A general wireless relay network.}\label{fig:randomrelays}
\end{centering}
\end{figure}

\begin{figure}[!htbp]
\centering
\includegraphics[width=0.4\textwidth]{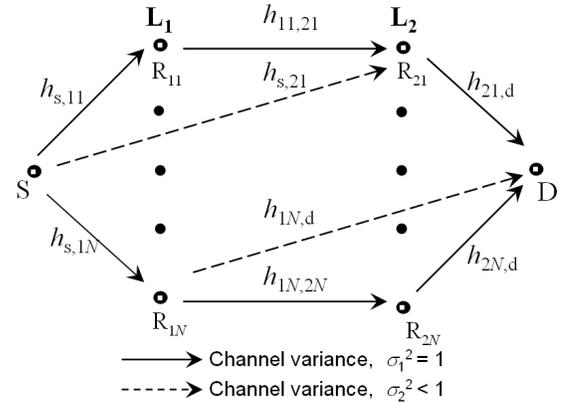}
\caption{System model.}\label{fig:sysmodel}
\end{figure}

A general wireless relay network is depicted in Fig.~\ref{fig:randomrelays}.  One source (S), one destination (D), and 2$N$ relays constitute this network.  Let us assume that paths from source to $N$ of these relays have low power loss and that to rest of the $N$ relays have higher loss.  
Assume that transmission is carried out in three different phases.
Then these can be grouped into two layers, having $N$ relays each, as shown in Fig.~\ref{fig:sysmodel}.
Here firm lines indicate stronger paths with identical low power loss while the dashed lines indicate weaker paths having equal high power loss. \\
\indent Let us introduce the notations used in this paper now.  L$_1$ and $\text{L}_2$   denote the first and the second layers as shown in Fig.~\ref{fig:sysmodel}.  $\text{R}_{ij}$ is the $j$th relay in the $i$th layer. The channel coefficients are designated $h_{s,1j}, h_{1j,2l}$ and $h_{2l,d}$ for S to $\text{R}_{1j}, \text{R}_{1j}$ to $\text{R}_{2l}$, and $\text{R}_{2l}$ to D respectively.  Superscript $k$, if used in channel coefficients, denotes the phase.\\
\indent For a complex matrix $\mathbf{A}, \left|\mathbf{A}\right|, \mathbf{A}^\text{H}, \mathbf{A}^\text{T}$ and $\mathbf{A}^*$ denote determinant, Hermitian, transpose and conjugate of $\mathbf{A}$ respectively.  $\mathbf{I}_T$ denotes the $T \times T$ identity matrix. For a vector $\mathbf{a}, \left\|\mathbf{a}\right\|$ denotes the norm of $\mathbf{a}$.  $\left\lfloor\delta\right\rfloor$ denotes the biggest integer smaller than or equal to $\delta$. \\
\indent It is assumed that channels are Rayleigh fading and quasi static with a coherence interval of at least $T$ symbol duration.\\
\indent  The scheme proposed by Jing and Hassibi (\cite{jing}) considered only S, L$_1$, and D in Fig. \ref{fig:sysmodel}.  The channel variance from S to L$_1$ and L$_1$ to D was assumed to be constant at unity by the authors.  The scheme consisted of two phases; in phase 1, S transmits and in phase 2, L$_1$ layer relays encode their received signals using a matrix of their own and transmit to D.  The authors proved that this effectively obtains a DSTC and achieves the same diversity as that of a multiple-antenna system with little degradation.  Let us call this basic protocol as Jing Hassibi Scheme (JHS).  Now let us prepare to derive different protocols from our model shown in Fig. \ref{fig:sysmodel} based on JHS.\\
\indent Assume that $\mathbf{s}=[s(1)\cdots s(T)]^{\text{T}}$ is the transmitted signal from S during a block of length $T$, when the channel coefficients are assumed to remain constant.  Assume also that the signal $\mathbf{s} \in\Omega= \{\mathbf{s}_1,\cdots,\mathbf{s}_L\} \subset \mathbb{C}^{T \times 1}$ is selected from $\Omega$, whose cardinality is $L$, for transmission and that $\mathbf{s}$ is normalized with $E[\mathbf{s}^\text{H}\mathbf{s}] = 1.$
Let $\mathbf{r}_{ij}^{(k)}$ denote the vector received by the relay $\text{R}_{ij}$ in phase $k$,
 $\mathbf{t}_{ij}^{(k)}$ denote the vector transmitted by $\text{R}_{ij}$ in phase $k$ multiplied with a factor and  $\mathbf{r}_{d}^{(k)}$ denote the received vector at destination in phase $k$ in a block duration $T$.\\
\indent Let $\sigma_1^{2}$ be the variance corresponding to the channel coefficients, $h_{s,1j}, h_{1j,2l}$, $h_{2l,d}$  and $\sigma_2^{2}$ be the variance corresponding to the channel coefficients $h_{s,2j}, h_{1j,d}$ for $1 \leq j,l \leq N.$  i.e. $E[\left|h_{s,1j}\right|^2]=E[\left|h_{1j,2l}\right|^2]=E[\left|h_{2l,d}\right|^2]=\sigma^{2}_1$ and  $E[\left|h_{s,2j}\right|^2]=E[\left|h_{1j,d}\right|^2]=\sigma^{2}_2$. $\sigma^{2}_1 > \sigma^{2}_2$ as discussed earlier and also assume with no loss of generality that $\sigma_1^2=1$.\\  
Assume that $\mathbf{u}_{ij}^{(k)}$ and $\mathbf{u}_{d}^{(k)}$ are the noise vectors added at the relay $\text{R}_{ij}$ and the destination, D respectively during the $k$th phase.  Let the components of these vectors be zero--mean white Gaussian independent random variables with variance $\sigma^{2}_n$.  By keeping $\sigma^{2}_n$ = 1 throughout, SNR is varied by  varying $P$, the total average power per symbol duration of the system.\\
\indent Each of the relays,  $\text{R}_{ij}$, have their own matrices, $\mathbf{A}_{ij}$, given by 
\begin{equation}
\mathbf{A}_{ij} =  \left[a_{ij}^{kl} 
\right] \label{Aij}
\end{equation}
which they use to finally produce a distributed space-time code \cite{jing}.  Here $k$ and $l$ denote the row and column numbers respectively.   These matrices are random real orthogonal with $\mathbf{A}_{ij}\mathbf{A}_{ij}^{\text{T}} = \mathbf{I}_T$ and each of the components, $a_{ij}^{kl}$ is zero mean Gaussian independent random variable with variance $1/T$.  The performance of the system, in fact, has been proved to be the same, using simulations in Chapter \ref{simulations}, with real orthogonal instead of complex unitary matrices considered in \cite{jing}.
The vector notations used are defined below:
\begin{align*}
\mathbf{r}_{ij}^{(k)}=
\begin{bmatrix}
r_{ij}^{(k)}(1) \\
\vdots\\
r_{ij}^{(k)}(T)
\end{bmatrix}, \;
\mathbf{t}_{ij}^{(k)}=
\begin{bmatrix}
t_{ij}^{(k)}(1) \\
\vdots\\
t_{ij}^{(k)}(T)
\end{bmatrix}, \;
\mathbf{r}_{d}^{(k)}=
\begin{bmatrix}
r_{d}^{(k)}(1) \\
\vdots\\
r_{d}^{(k)}(T)
\end{bmatrix},
\end{align*}
\begin{align*}
\mathbf{u}_{ij}^{(k)}=
\begin{bmatrix}
u_{ij}^{(k)}(1) \\
\vdots\\
u_{ij}^{(k)}(T)
\end{bmatrix}, \text{ and }\;
\mathbf{u}_{d}^{(k)}=
\begin{bmatrix}
u_{d}^{(k)}(1) \\
\vdots\\
u_{d}^{(k)}(T)
\end{bmatrix}.
\end{align*}
In the next chapter we will derive different protocols from JHS suggested by Jing and Hassibi \cite{jing}.

%% file: sec_protocols.tex
\section{Protocols derived from JHS} \label{protocolsfromJHS}
Five protocols have been derived from the one proposed in \cite{jing}.  Let us assume that all these protocols operate in three phases of $T$ symbol duration each, with an available total average power of $PT$. As the first phase is the same for all the protocols, we will see it here and see the second and third phases in corresponding sections.\\  
\indent Refer to the System Model discussed in Chapter \ref{systemmodel} shown in Fig.~\ref{fig:sysmodel}.  In phase 1, S transmits $c_1s(\tau)$ at time $\tau$, for $1\leq\tau\leq T$.  i.e. S transmits $c_1\mathbf{s}$ during $T$ symbol duration, where $\mathbf{s}=[s(1)\cdots s(T)]^{\text{T}}$.  $R_{ij}$ receives $r_{ij}^{(1)}(\tau)=c_1s(\tau)h_{s,ij}+u_{ij}^{(1)}(\tau)$ at time $\tau$ and in vector form
\begin{equation}
\mathbf{r}_{ij}^{(1)}=c_1\mathbf{s}h_{s,ij}+\mathbf{u}_{ij}^{(1)}.
\label{rij1}
\end{equation}
To find $c_1$ the power transmitted by S is to be known. If we assume that $p_1$ is the power transmitted per symbol duration by S, then
\begin{align}
p_1T=E(c_1\mathbf{s}^\text{H}\mathbf{s}c_1)=c_1^2 \Rightarrow c_1=\sqrt{p_1T}. \label{c1}
\end{align}
The average power received by $R_{1j}$ in $T$ symbol duration is
\begin{align}
E[\mathbf{r}_{1j}^{(1)\text{H}}\mathbf{r}_{1j}^{(1)}] =& E[(c_1\mathbf{s}^\text{H}h_{s,1j}^*+\mathbf{u}_{1j}^{(1)\text{H}})(c_1\mathbf{s}h_{s,1j}+\mathbf{u}^{(1)}_{1j})]\nonumber \\
=&c_1^2E[|h_{s,1j}|^2]+E[\mathbf{u}_{1j}^{(1)\text{H}}\mathbf{u}_{1j}^{(1)}]\nonumber \\=&c_1^2+T.\label{powerreceR1jphase1MJHS}
\end{align}
Equation \eqref{powerreceR1jphase1MJHS} is arrived at with the assumption that signal, noise, and channel are uncorrelated amongst each other with zero mean. Similarly it can be proved that the power received by $R_{2j}$ is
\begin{align}
E[\mathbf{r}_{2j}^{(1)\text{H}}\mathbf{r}_{2j}^{(1)}] = \sigma_2^2c_1^2+T. \label{powreceR2jMJHS}
\end{align}
Let us see a detailed description of each one of the five derived protocols, while also discussing their second and third phases, in the following sections.
\subsection{Relay Matrix Combining (RMC)} \label{RMCintro}
\begin{figure}[!tb]
\centering
\includegraphics[width=0.4 \textwidth]{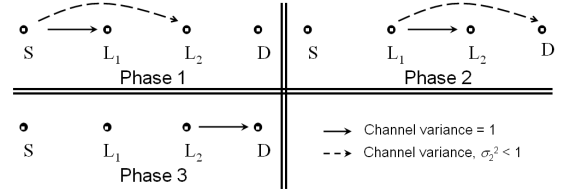}
\caption{Various phases in RMC/RSC/RMCKC.}\label{fig:RMCRSCRMCKCphases}
\end{figure}

Different phases of transmission and reception of this protocol are shown in Fig.~\ref{fig:RMCRSCRMCKCphases} and explained below:   
\begin{itemize}
\item Phase 1: S transmits; L$_1$ and L$_2$ layer relays receive.
\item Phase 2: L$_1$ layer relays transmit; L$_2$ layer relays and D receive.
\item Phase 3: L$_2$ layer relays transmit and D receives.
\end{itemize}
\indent As the name suggests, this system combines the two vectors received by L$_2$ in phases 1 and 2, using a matrix before transmission in third phase.  Let $p_2/N$ and $p_3/N$ be the power transmitted per symbol duration by each of the relays in the second and third phases respectively.
\subsubsection{Protocol Analysis}
In phase 2 the relays $\text{R}_{1j}, 1 \leq j \leq N,$ transmit $c_2t_{1j}^{(2)}(\tau)$ at time $\tau$ for $1\leq \tau \leq T $ where $t_{1j}^{(2)}(\tau)=\sum_{p=1}^Ta_{1j}^{\tau p}r_{1j}^{(1)}(p)$ and in vector form
\begin{equation}
\mathbf{t}_{1j}^{(2)}=\mathbf{A}_{1j}\mathbf{r}_{1j}^{(1)}
\label{t1j2RMC}
\end{equation}
where $\mathbf{A}_{1j}$ is shown in \eqref{Aij} with $i=1$.
The relays in $\text{L}_2$ receive $\mathbf{r}_{2j}^{(2)}$ and D receives $\mathbf{r}_{d}^{(2)}.$  These can be proved to be
\begin{align}
\mathbf{r}_{2j}^{(2)}= \mathbf{S}_1\mathbf{h}_{s,1,2j}+ c_2\sum_{i=1}^Nh_{1i,2j}\mathbf{A}_{1i}\mathbf{u}_{1i}^{(1)} +\mathbf{u}_{2j}^{(2)}
\label{r2j2RMC}
\end{align} 
and
\begin{align}
\mathbf{r}_{d}^{(2)}=\mathbf{S}_1\mathbf{h}_{s,1,d}+\mathbf{u}_{x} \label{rd2RMC}
\end{align} 
where \\
\begin{align*}
\mathbf{S}_1=&c_1c_2[\mathbf{A}_{11}\mathbf{s} \ldots \mathbf{A}_{1N}\mathbf{s}],~~~
\mathbf{h}_{s,1,2j}=
\begin{bmatrix}
h_{s,11}h_{11,2j} \\
\vdots\\
h_{s,1N}h_{1N,2j}
\end{bmatrix},
\end{align*}
\begin{align}
\mathbf{h}_{s,1,d}=&
\begin{bmatrix}
h_{s,11}h_{11,d} \\
\vdots\\
h_{s,1N}h_{1N,d}
\end{bmatrix},\text{ and } \nonumber \\
\mathbf{u}_{x}=&c_2\sum_{i=1}^Nh_{1i,d}\mathbf{A}_{1i}\mathbf{u}_{1i}^{(1)}+\mathbf{u}_d^{(2)}. \label{uxRMC}
\end{align}
Like in JHS \cite{jing}, it has been proved in \eqref{rd2RMC} that the distributed space-time code in this case is $\mathbf{S}_1$ and the equivalent channel matrix is $\mathbf{h}_{s,1,d}$ with the equivalent noise vector $\mathbf{u}_{x}$.
To find $c_2$ we require to get an expression for the power transmitted by each relay, $p_2T/N$, which is $E\left[c_2\mathbf{t}_{ij}^{(2)\text{H}}\mathbf{t}_{ij}^{(2)}c_2\right].$  The available power $p_2T$ is equally divided amongst $N$ relays as the variance of the channel coefficients are the same for all of them.  The power transmitted by each relay can be proved to be $c_2^2(p_1+1)T$, which leads to
\begin{align}
c_2=\sqrt{\frac{p_2}{N(p_1+1)}}. \label{c2RMC}
\end{align}
\indent In phase 3, the two received vectors $\mathbf{r}_{2j}^{(1)}$ and $\mathbf{r}_{2j}^{(2)}$ are transmitted, by R$_{2j}$, after a matrix combining operation on the stacked vector
\begin{align}
\mathbf{r}_{2j}=
\begin{bmatrix}
\mathbf{r}_{2j}^{(1)} \\
\mathbf{r}_{2j}^{(2)} \\
\end{bmatrix}, \label{r2jRMC}
\end{align}
namely
$\mathbf{t}_{2j}^{(3)}=\mathbf{A}_{2j}^{'}\mathbf{r}_{2j}.$
The matrix $\mathbf{A}_{2j}^{'}$ (size $T \times 2T$) is the relay matrix of R$_{2j}$ and is also orthogonal like its counterpart in L$_1$ relays, and given by 
\begin{equation*}
\mathbf{A}_{2j}^{'}=\frac{1}{\sqrt{2}}\left[\begin{array}{ccc}
a_{2j}^{11} & \cdots & a_{2j}^{1,2T}\\
\vdots      & \ddots & \vdots \\
a_{2j}^{T1} & \cdots & a_{2j}^{T,2T}
\end{array}\right].
\end{equation*}
$\mathbf{A}_{2j}^{'}$'s can also be written in the submatrix form as
\begin{align}
\mathbf{A}_{2j}^{'}=\frac{1}{\sqrt{2}}\left[\begin{array}{ccc}
\mathbf{A}_{2j}(1)&|&\mathbf{A}_{2j}(2)
\end{array}\right] \label{A2jprimeRMC}
\end{align}
where $\mathbf{A}_{2j}(1)$ and $\mathbf{A}_{2j}(2)$ are the submatrices of $\mathbf{A}_{2j}^{'}$ with first $T$ columns and the last $T$ columns respectively.  Also these submatrices are chosen to be orthogonal.\\
\indent Hence the vector transmitted by R$_{2j}$ is $c_3\mathbf{t}_{2j}^{(3)}$ and therefore the vector received by D is $ \mathbf{r}_d^{(3)} $ where the components are given by 
\begin{align*}
r_d^{(3)}(\tau)=\sum_{i=1}^N c_3t_{2i}^{(2)}(\tau)h_{2i,d}+u_d^{(3)}(\tau).
\end{align*}
It can be proved after some calculations that
\begin{align}
\mathbf{r}_d^{(3)}=&\frac{c_1c_3}{\sqrt{2}}\mathbf{S}_2(1)\mathbf{h}_{s,2,d}+ \mathbf{u}_{z}\nonumber \\  
+&\frac{c_1c_2c_3}{\sqrt{2}}[\mathbf{S}_{21,1}(2)\mathbf{h}_{s,1,21} \ldots
\mathbf{S}_{2N,1}(2)\mathbf{h}_{s,1,2N}]\mathbf{h}_{2,d}
\label{rd3RMC}
\end{align}
where 
\begin{align}
\mathbf{S}_2(1)=\left[\mathbf{A}_{21}(1)\mathbf{s} \ldots \mathbf{A}_{2N}(1)\mathbf{s}\right],~~ \mathbf{h}_{s,2,d}=
\begin{bmatrix}
h_{s,21}h_{21,d} \\
\vdots\\
h_{s,2N}h_{2N,d}
\end{bmatrix},
\end{align}
\begin{align}
\mathbf{S}_{2n,1}(2)=[\mathbf{A}_{2n}(2)\mathbf{A}_{11}\mathbf{s} \ldots \mathbf{A}_{2n}(2)\mathbf{A}_{1N}\mathbf{s}],
\end{align}
\begin{align}
\mathbf{h}_{s,1,2n}=
\begin{bmatrix}
h_{s,11}h_{11,2n} \\
\vdots \\
h_{s,1N}h_{1N,2n}
\end{bmatrix},~~
\mathbf{h}_{2,d}=
\begin{bmatrix}
h_{21,d} \\
\vdots \\
h_{2N,d}
\end{bmatrix}, \label{h2dRMC}
\end{align}
and
\begin{align}
\mathbf{u}_{z}=&\frac{c_3}{\sqrt{2}}\sum_{j=1}^N\left[\mathbf{A}_{2j}(1)\mathbf{u}_{2j}^{(1)}+\mathbf{A}_{2j}(2)\mathbf{u}_{2j}^{(2)}\right]h_{2j,d} \nonumber \\ 
+&\frac{c_2c_3}{\sqrt{2}}\sum_{i=1}^N\sum_{j=1}^Nh_{1i,2j}h_{2j,d}\mathbf{A}_{2j}(2)\mathbf{A}_{1i}\mathbf{u}_{1i}^{(1)}+\mathbf{u}_d^{(3)}. \label{uzRMC}
\end{align}
Here $\mathbf{A}_{2j}(l), l=1,2$ are given in \eqref{A2jprimeRMC} and $1 \leq n \leq N$ in \eqref{h2dRMC}.  To find $c_3$ let us find the power transmitted by each relay in L$_2$.  This is given by $p_3T/N=E\left(c_3\mathbf{t}_{2j}^{(3)\text{H}}\mathbf{t}_{2j}^{(3)}c_3\right).$  The total available power $p_2T$ is equally divided amongst $N$ relays as the variance of the channel coefficients are same for all of them.  The power transmitted by each relay can be proved to be $c_3^2T\left[2+p_1\sigma_2^2+p_2\right]$, which leads to
\begin{align}
c_3=\sqrt{\frac{p_3}{N(2+p_1\sigma_2^2+p_2)}}.\label{c3RMC}
\end{align}
\indent All the transmission vectors and the multiplication factors are summarized in Table \ref{tableRMC}.  It can be seen from \eqref{rd3RMC} that the space-time code here has been mingled up with the channel.  Nevertheless an ML decoder has been derived for this protocol.
\begin{table}[htb]
\begin{center}
\caption{Transmitted vectors and multiplication factors - RMC} \label{tableRMC}
\vspace{0.2cm}
\begin{tabular}{|c|c|c|}\hline
\tt{Vector}&\tt{Factor}&\tt{Transmitted by}\\ \hline\hline
$\mathbf{s}$&$c_1=\sqrt{p_1T}$&S in phase 1\\ \hline
$\mathbf{t}^{(2)}_{1j}=\mathbf{A}_{1j}\mathbf{r}^{(1)}_{1j}$&$c_2=\sqrt{\frac{p_2}{N(p_1+1)}}$&$\text{L}_{1}$ relays in phase 2\\ \hline
$\mathbf{t}^{(3)}_{2j}=\mathbf{A}_{2j}^{'}\mathbf{r}_{2j}$&$c_3=\sqrt{\frac{p_3}{N(2+p_1\sigma_2^2+p_2)}}$&$\text{L}_{2}$ relays in phase 3 \\\hline \end{tabular}
\end{center}
\end{table}
\subsubsection{ML Decoder}
D has two received vectors namely, $\mathbf{r}_d^{(2)}=\mathbf{x}$, say and $\mathbf{r}_d^{(3)}=\mathbf{z}$, say as shown in \eqref{rd2RMC} and \eqref{rd3RMC} respectively.  
These two vectors are stacked as 
\begin{align}
\mathbf{y}=
\begin{bmatrix}
\mathbf{x}\\
\mathbf{z}
\end{bmatrix}.
\end{align}
The likelihood function that $\mathbf{s}$ is transmitted is $\Pr(\mathbf{y}|\mathbf{s})$.  To find an expression for this (given in \eqref{Prys}) we have to know the nature of the joint density function.  Let us first consider $\mathbf{x}$ and $\mathbf{z}$ separately.  It can be seen from \eqref{rd2RMC} that $\mathbf{x}$ is jointly Gaussian and from \eqref{rd3RMC} that $\mathbf{z}$ is jointly Gaussian.  Also the mean of $\mathbf{x}|\mathbf{s}$ is
\begin{align}
E[\mathbf{x}|\mathbf{s}]=c_1c_2[\mathbf{A}_{11}\mathbf{s} \ldots \mathbf{A}_{1N}\mathbf{s}]\mathbf{h}_{s,1,d}=\mathbf{m}_x,\text{ say}
\label{mxRMC}
\end{align}
and the covariance matrix of $\mathbf{x}|\mathbf{s}$ can be worked out to be  
\begin{align}
E\left[(\mathbf{x}-\mathbf{m}_x)(\mathbf{x}-\mathbf{m}_x)^\text{H}
|\mathbf{s}\right]=&\left(1+c_2^2\sum_{j=1}^N|h_{1j,d}|^2\right)\mathbf{I}_T \nonumber \\
=&\mathbf{P}_x, \text{ say.}
\label{PxRMC}
\end{align}
Similarly we can obtain $\mathbf{m}_z$ and $\mathbf{P}_z$ as
\begin{align}
\mathbf{m}_z=E[\mathbf{z}|\mathbf{s}]=\frac{c_1c_3}{\sqrt{2}} \mathbf{S}_2(1) \mathbf{h}_{s,2,d}+\frac{c_1c_2c_3}{\sqrt{2}}\mathbf{S}_{22h}\mathbf{h}_{2,d}
\label{mzRMC}
\end{align}
and
\begin{align}
\mathbf{P}_z =&E\left[\left(\mathbf{z}-\mathbf{u}_z\right)\left(\mathbf{z}-\mathbf{u}_z\right)^\text{H}\right]=\left[1+c_3^2\sum_{j=1}^N|h_{2j,d}|^2\right]\mathbf{I}_T \nonumber \\
+&\frac{c_2^2c_3^2}{2}\sum_{i=1}^N\sum_{j=1}^N\sum_{k=1}^Nh_{1i,2j}h_{2j,d}h_{1i,2k}^*h_{2k,d}^*\textbf{A}_{2j}(2)\textbf{A}_{2k}^\text{T}(2). \label{PzRMC}
\end{align}
From the above discussions we can see that $\mathbf{y}$ is also jointly Gaussian with mean vector and covariance matrix given by \cite{shalom}
\begin{align}
\mathbf{m}_y=
\begin{bmatrix}
\mathbf{m}_x \\
\mathbf{m}_z
\end{bmatrix}\text{ and } \;
\mathbf{P}_y=
\begin{bmatrix}
\mathbf{P}_x & \mathbf{P}_{xz}\\
\mathbf{P}_{zx} & \mathbf{P}_z\\
\end{bmatrix} \label{meancovarianceyRMC}
\end{align}
respectively.  Here $\mathbf{P}_{xz}$ and $\mathbf{P}_{zx}$ are the cross covariance matrices given by 
\begin{align*}
\mathbf{P}_{xz}=&E\left[(\mathbf{x}-\mathbf{m}_x)(\mathbf{z}-\mathbf{m}_z)^\text{H}|\mathbf{s}\right]=\mathbf{P}_{zx}^\text{H}
\end{align*} 
and we can derive
\begin{align}
\mathbf{P}_{xz}=&\frac{c_2^2c_3}{\sqrt{2}}\sum_{i=1}^N\sum_{j=1}^Nh_{1i,d}h_{1i,2j}^*h_{2j,d}^*\mathbf{A}_{2j}^\text{T}(2).
\end{align}
Now as $\mathbf{y}$ is complex Gaussian we can write \cite{nielsen}
\begin{align}
\Pr(\mathbf{y}|\mathbf{s})=\frac{\exp\left[{-(\mathbf{y}-\mathbf{m}_y)^\text{H}\mathbf{P}_y^{-1}(\mathbf{y}-\mathbf{m}_y)}\right]}{\pi^{2T}|\mathbf{P}_y|} \label{Prys}
\end{align}
where $\mathbf{m}_y$ and $\mathbf{P}_y$ are given in \eqref{meancovarianceyRMC}.
Hence we can write the decoded vector as \cite{shalom}
\begin{align}
\widehat{\mathbf{s}}=&\arg\max_{\mathbf{s}}\Pr(\mathbf{y}|\mathbf{s})=\arg\min_{\mathbf{s}}\left\|\mathbf{y}'\right\|^2
\end{align}
where
\begin{align*}
\mathbf{y}'=&\mathbf{P}_y^{-\frac{1}{2}}(\mathbf{y}-\mathbf{m}_y).
\end{align*}
\subsubsection{Receive SNR} \label{receiveSNRRMC}
Let us derive an expression for receive SNR. We have two received signal vectors at the destination namely, $\mathbf{r}_d^{(2)}$ shown in \eqref{rd2RMC} and $\mathbf{r}_d^{(3)}$ shown in \eqref{rd3RMC}.  The received signal power and noise power in second phase can be written as $P_s^{(2)}=E[\mathbf{m}_x^{\text{H}}\mathbf{m}_x]$ and $P_n^{(2)}=E[\mathbf{u}_{x}^{\text{H}}\mathbf{u}_{x}]$ respectively.  Hence from \eqref{uxRMC} and \eqref{mxRMC}
\begin{align*}
P_s^{(2)}=E\left[c_1^2c_2^2\sum_{j=1}^N\sum_{i=1}^Nh_{1j,d}^*h_{s,1j}^*\mathbf{s}^{\text{H}}\mathbf{A}_{ij}^{\text{H}}\mathbf{A}_{1i}\mathbf{s}h_{s,1i}h_{1i,d}\right] \end{align*}
and
\begin{align*}
P_n^{(2)}=&E\left[c_2^2\sum_{j=1}^N\sum_{i=1}^Nh_{1j,d}^*\mathbf{u}_{1j}^{(1)\text{H}}\mathbf{A}_{ij}^{\text{H}}\mathbf{A}_{1i}\mathbf{u}_{1i}^{(1)}h_{1i,d}\right]\nonumber \\
+&E\left[\mathbf{u}_d^{(2)\text{H}}\mathbf{u}_d^{(2)}\right].
\end{align*}
Now as the channel coefficients are all independent and zero mean, unless $i=j$, the expected values will be zero.  So the above equations simplify to,
\begin{align}
P_s^{(2)}=&c_1^2c_2^2\sum_{j=1}^NE\left[\left|h_{1j,d}\right|^2\right]E\left[\left|h_{s,1j}\right|^2\right]E\left[\mathbf{s}^{\text{H}}\mathbf{A}_{ij}^{\text{H}}\mathbf{A}_{1j}\mathbf{s}\right] \nonumber \\
=&c_1^2c_2^2N\sigma_2^2 \text{ and} \nonumber \\
P_n^{(2)}=&c_2^2\sum_{j=1}^NE\left[\left|h_{1j,d}\right|^2\right]E\left[\mathbf{u}_{1j}^{\text{H}}\mathbf{A}_{ij}^{\text{H}}\mathbf{A}_{1j}\mathbf{u}_{1j}\right]+E\left[\mathbf{u}_d^{(2)\text{H}}\mathbf{u}_d^{(2)}\right] \nonumber \\
=&c_2^2TN\sigma_2^2+T.
\end{align}
Similarly, $P_s^{(3)}$ and $P_n^{(3)}$ can be derived from \eqref{rd3RMC} as 
\begin{align}
P_s^{(3)}=&\frac{1}{2}\left[c_1^2c_3^2N\sigma_2^2+c_1^2c_2^2c_3^2N^2\right] \text { and} \nonumber \\
P_n^{(3)}=&c_3^2TN+\frac{c_2^2c_3^2TN^2}{2}+T.
\end{align}
The receive SNR is then
\begin{align*}
\text{snr}_{\text{RMC}}=&\frac{P_s^{(2)}+P_s^{(3)}}{P_n^{(2)}+P_n^{(3)}} \\
=&\frac{2c_1^2c_2^2N\sigma_2^2+c_1^2c_3^2N\sigma_2^2+c_1^2c_2^2c_3^2N^2}{2c_2^2NT\sigma_2^2+2c_3^2NT+c_2^2c_3^2N^2T+4T}.
\end{align*}
Substituting the values of $c_1,~c_2,$ and $c_3$ we can obtain equation \eqref{snrRMC} shown at the top of next page.  
\begin{figure*}[!t]
\normalsize
\begin{eqnarray}
\text{snr}_{\text{RMC}}& = & \frac{p_1\left[2p_2^2\sigma_2^2+(1+p_1)p_3\sigma_2^2+p_2(p_3+4\sigma_2^2+2p_1\sigma_2^4)\right]}{\begin{array}{c}4p_1^2\sigma_2^2+(2+p_2)(4 + p_3 + 2p_2\sigma_2^2+2p_1(4+p_3+2\sigma_2^2+p_2(2 +\sigma_2^4))\end{array}}.
\label{snrRMC}
\end{eqnarray}
\hrulefill
\vspace*{4pt}
\end{figure*}
Now allocation of $p_1, p_2,$ and $p_3$ can be done by maximizing the receive SNR shown in \eqref{snrRMC}.  But as it is quite tedious a fine computer search is resorted to as discussed in Section \ref{optpowalloc}.
\subsection{Extended Jing Hassibi Scheme (EJHS)} \label{EJHSintro}
\begin{figure}[!tb]
\centering
\includegraphics[width=0.4 \textwidth]{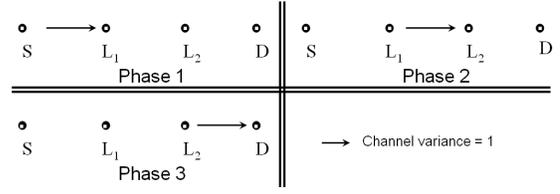}
\caption{Various phases in EJHS.}\label{fig:EJHSphases}
\end{figure}

This is named so, as the JHS suggested by \cite{jing} has been extended here to have one extra layer.  The derivation and analysis of this simple protocol is warranted as the performance of EJHS forms a base line for comparison with other derived protocols.\\
\indent Different phases of transmission and reception in this protocol are shown in Fig. \ref{fig:EJHSphases} and explained below:
\begin{itemize}
\item Phase 1: S transmits; L$_1$ layer relays receive.
\item Phase 2: L$_1$ layer relays transmit and L$_2$ layer relays receive.
\item Phase 3: L$_2$ layer relays transmit and D receives.
\end{itemize}
\subsubsection{Protocol Analysis}
Phase 2 is exactly similar to that of RMC, except that D neglects any signal received.  Let $p_2/N$ be the average power transmitted per symbol duration by each of the $\text{L}_1$ relays in this phase.  Hence $\mathbf{r}_{2j}^{(2)}$ is the same as that of RMC and is given by equation \eqref{r2j2RMC}.\\
\indent In phase 3, let $p_3/N$ be the average power transmitted per symbol duration by each of the $\text{L}_2$ relays.  The vector that is transmitted by $\text{R}_{2j}$ is
\begin{align}
c_3\mathbf{t}_{2j}^{(3)}=c_3\mathbf{A}_{2j}\mathbf{r}_{2j}^{(2)}.
\label{t2j3EJHS}
\end{align}
The vector received by destination is $ \mathbf{r}_d^{(3)} $ where the components are given by 
\begin{align}
r_d^{(3)}(\tau)=\sum_{i=1}^N c_3t_{2i}^{(2)}(\tau)h_{2i,d}+u_d^{(3)}(\tau). \nonumber
\end{align}
It can be proved after some calculations that
\begin{align}
\mathbf{r}_d^{(3)}=\mathbf{m}_z+\mathbf{u}_{z}=\mathbf{z}, \text{ say}
\label{rd3EJHS}
\end{align}
\text{where }
\begin{align}
\mathbf{m}_z=&c_1c_2c_3[\mathbf{S}_{21}\mathbf{h}_{s,1,21} \ldots 
\mathbf{S}_{2N}\mathbf{h}_{s,1,2N}]\mathbf{h}_{2,d}, \label{mzEJHS}
\end{align}
and
\begin{align}
\mathbf{S}_{2n}=&[\mathbf{A}_{2n}\mathbf{A}_{11}\mathbf{s} \ldots \mathbf{A}_{2n}\mathbf{A}_{1N}\mathbf{s}],~~1 \leq n \leq N. \label{S2nEJHS}
\end{align}
Here $\mathbf{h}_{s,1,2n}$ and $\mathbf{h}_{2,d}$ are defined earlier in equation \eqref{h2dRMC}.
Also
\begin{align}
\mathbf{u}_{z}=&c_2c_3\sum_{i=1}^N\sum_{j=1}^Nh_{1i,2j}h_{2j,d}\mathbf{A}_{2j}\mathbf{A}_{1i}\mathbf{u}_{1i}^{(1)}\nonumber \\
+&c_3\sum_{j=1}^Nh_{2j,d}\mathbf{A}_{2j}\mathbf{u}_{2j}^{(2)}+\mathbf{u}_d^{(3)}. \label{uzEJHS}
\end{align}
$c_2$ is the same as that of RMC shown in \eqref{c2RMC}.  To find $c_3$ the power transmitted by each relay in L$_2$ is to be found out.  This is given by $p_3T/N=E\left(c_3\mathbf{t}_{2j}^{(3)\text{H}}\mathbf{t}_{2j}^{(3)}c_3\right).$  This can be proved to be $c_3^2T\left[1+p_2\right]$ which implies 
\begin{align}
c_3=\sqrt{\frac{p_3}{N(1+p_2)}}. \label{c3EJHS}
\end{align}
The transmission vectors and the corresponding multiplication factors are summarized in Table \ref{tableEJHS}.
\begin{table}[htb]
\begin{center}
\caption{Transmitted vectors and multiplication factors - EJHS} \label{tableEJHS}
\vspace{0.2cm}
\begin{tabular}{|c|c|c|}\hline
\tt{Vector}&\tt{Factor}&\tt{Transmitted by}\\ \hline\hline
$\mathbf{s}$&$c_1=\sqrt{p_1T}$&S in phase 1\\ \hline
$\mathbf{t}^{(2)}_{1j}=\mathbf{A}_{1j}\mathbf{r}_{1j}^{(1)}$&$c_2=\sqrt{\frac{p_2}{N(p_1+1)}}$&$\text{L}_{1}$ relays in phase 2\\ \hline
$\mathbf{t}^{(3)}_{2j}=\mathbf{A}_{2j}\mathbf{r}_{2j}^{(2)}$&$c_3=\sqrt{\frac{p_3}{N(1+p_2)}}$&$\text{L}_{2}$ relays in phase 3 \\ \hline
\end{tabular}
\end{center}
\end{table}
\subsubsection{ML Decoder}
Unlike in RMC case, EJHS has only one receive vector at D.  We can prove that this vector, $\mathbf{z}$, shown in \eqref{rd3EJHS}, is complex Gaussian with mean $\mathbf{m}_z$, shown in \eqref{mzEJHS}, and covariance matrix $\mathbf{P}_z$, where  $\mathbf{P}_z$ can be derived to be 
\begin{align}
\mathbf{P}_z=&\left[1+c_3^2\sum_{j=1}^N|h_{2j,d}|^2\right]\mathbf{I}_T\nonumber \\ +&c_2^2c_3^2\sum_{i=1}^N\sum_{j=1}^N\sum_{k=1}^N\mathbf{A}_{2i}\mathbf{A}_{2k}^Th_{2i,d}h_{2k,d}^*h_{1j,2i}h_{1j,2k}^*.
\end{align}
Now the decoded vector can be proved to be 
\begin{align}
\widehat{\mathbf{s}}=\arg\max_{\mathbf{s}}\Pr(\mathbf{z}|\mathbf{s})=\arg\min_{\mathbf{s}}\left\|\mathbf{z}'\right\|^2
\end{align}
where $\mathbf{z}'=\mathbf{P}_z^{-\frac{1}{2}}(\mathbf{z}-\mathbf{m}_z).$
\subsubsection{Receive SNR} \label{receiveSNREJHS}
On similar lines as was done in RMC in Section \ref{receiveSNRRMC}, we can prove that the receive SNR in this case is
\begin{align}
\text{snr}_{\text{EJHS}}=&\frac{c_1^2c_2^2c_3^2N^2}{c_2^2c_3^2N^2T+c_3^2NT+T}  \nonumber \\ 
=&\frac{p_1p_2p_3}{(1+p_2)(1+p_3)+p_1(1+p_2+p_3)}.\label{snrEJHS}
\end{align}
It can also be derived that $\text{snr}_{\text{EJHS}}$ attains the maximum value of 
\begin{align}
\frac{P^3}{9(3+3P+P^2)}
\end{align}
when $p_1=p_2=p_3=1/3$.  This has also been verified in Section \ref{optpowalloc} using simulations.  Hence in the BER simulations in Section \ref{BERplots} for EJHS, the total power is divided equally amongst the three phases accordingly.\\
\indent If $\sigma_2^2$ is very low ($< 0.01$), then EJHS is expected to perform better than all protocols as it neglects these weaker signals.  This is verified in Section \ref{BERplots} using simulations. 
\subsection{Modified Jing Hassibi Scheme (MJHS)}\label{MJHSintro}
\begin{figure}[!tb]
\centering
\includegraphics[width=0.4 \textwidth]{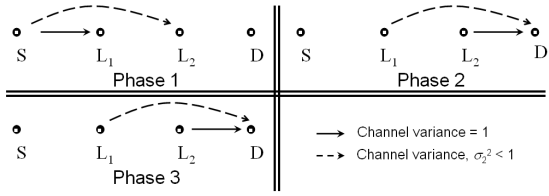}
\caption{Various phases in MJHS.}\label{fig:MJHSphases}
\end{figure}

As the name suggests the JHS has been modified in this protocol.
Different phases of transmission and reception in MJHS case are shown in Fig.~\ref{fig:MJHSphases} and explained below:
\begin{itemize}
\item Phase 1: S transmits; L$_1$ and L$_2$ layer relays receive.
\item Phase 2: L$_1$ layer and L$_2$ layer relays transmit;  and D receives.
\item Phase 3: L$_1$ layer and L$_2$ layer relays transmit;  and D receives.
\end{itemize}
\subsubsection{Protocol Analysis} \label{protocolanalysisMJHS}
In this protocol, we have phase 3 exactly similar to phase 2 so as to keep the total time duration to be 3$T$, similar to the other protocols.
Let $p_2T/2$ be the power transmitted by L$_1$ and $p_3T/2$ by L$_2$ relays in the second phase.  As the vectors to be transmitted by L$_1$ and L$_2$ relays in the second and third phases are identical and that the channel is assumed to have the same statistics, we have equally divided the power between the second and third phases.
Let $c_{21}\mathbf{t}_{1j}^{(k)}$ and $c_{22}\mathbf{t}_{2j}^{(k)}$ be the vectors transmitted by R$_{1j}$ and R$_{2j}$ relays respectively, in $k$th phase, with $k=2,3$.  Average power transmitted by R$_{1j}$ in $T$ channel uses during the $k$th phase is
\begin{align}
E[c_{21}\mathbf{t}_{1j}^{(k)\text{H}}c_{21}\mathbf{t}_{1j}^{(k)}] =& c_{21}^2E[\mathbf{t}_{1j}^{(k)\text{H}}\mathbf{t}_{1j}^{(k)}]\nonumber \\
=&c_{21}^2E[(\mathbf{r}_{1j}^{(1)\text{H}}\mathbf{A}_{1j}^\text{H})(\mathbf{A}_{1j}\mathbf{r}_{1j}^{(1)})]\nonumber \\ =&c_{21}^2(c_1^2+T). \label{powertranR1jphase23MJHS}
\end{align}
Equation \eqref{powertranR1jphase23MJHS} is arrived from \eqref{powerreceR1jphase1MJHS} and $\mathbf{A}_{1j}^\text{H}\mathbf{A}_{1j}=\mathbf{I}_T$.
Hence total power transmitted by R$_{1j}$ alone in phase $k$, with $k=2,3$ is
\begin{align}
\frac{p_2T}{2N}=c_{21}^2(c_1^2+T) \nonumber \\ 
\Rightarrow c_{21}=\sqrt{\frac{p_2}{2N(1+p_1)}}.\label{c21MJHS} \end{align}
Here $c_1$ is substituted from \eqref{c1}.  Similarly it can be proved that the power transmitted by $R_{2j}$ in $T$ channel uses is
\begin{align}
E[c_{22}\mathbf{t}_{2j}^{(k)\text{H}}c_{22}\mathbf{t}_{2j}^{(k)}] = c_{22}^2(\sigma_2^2c_1^2+T). \label{powertranR2jMJHS}
\end{align}
Hence total power transmitted by R$_{2j}$ alone in phase $k$, with $k=2,3$ can be worked out to be
\begin{align}
\frac{p_3}{2N}=c_{22}^2(1+\sigma_2^2p_1) \nonumber \\ 
\Rightarrow c_{22}=\sqrt{\frac{p_3}{2N(1+\sigma_2^2p_1)}}.\label{c22MJHS}
\end{align}
It can be proved that the received vector at D in phase $k$ is
\begin{align}
\mathbf{r}_d^{(k)}=c_1\mathbf{S}_1\mathbf{h}_{s,d}+\mathbf{w}_{dk},
k=2,3, \label{rdkMJHS}
\end{align}
where $\mathbf{S}_1=\left[c_{21}\mathbf{A}_{11}\mathbf{s} \ldots c_{21}\mathbf{A}_{1N}\mathbf{s} c_{22}\mathbf{A}_{21}\mathbf{s} \ldots c_{22}\mathbf{A}_{2N}\mathbf{s}\right],$
\begin{align*}
\mathbf{h}_{s,d}=
\begin{bmatrix}
h_{s,11}h_{11,d} \\
\vdots \\
h_{s,1N}h_{1N,d} \\
h_{s,21}h_{21,d} \\
\vdots \\
h_{s,2N}h_{2N,d} \\
\end{bmatrix}, \;
\end{align*}
and
\begin{align}
\mathbf{w}_{dk}=\sum_{j=1}^Nc_{21}\mathbf{A}_{1j}\mathbf{u}_{1j}^{(1)}\mathbf{h}_{1j,d} +\sum_{j=1}^Nc_{22}\mathbf{A}_{2j}\mathbf{u}_{2j}^{(1)}\mathbf{h}_{2j,d}+\mathbf{u}_d^{(k)}. \label{wdkMJHS}
\end{align}
The transmission vectors and the corresponding multiplication factors are summarized in Table \ref{tableMJHS}.
\begin{table}[htb]
\begin{center}
\caption{Transmitted vectors and multiplication factors - MJHS} \label{tableMJHS}
\vspace{0.2cm}
\begin{tabular}{|c|c|c|}\hline
\tt{Vector}&\tt{Factor}&\tt{Transmitted by}\\ \hline\hline
$\mathbf{s}$&$c_1=\sqrt{p_1T}$&S in phase 1\\ \hline
$\mathbf{t}^{(2)}_{1j}=\mathbf{A}_{1j}\mathbf{r}_{1j}^{(1)}$&$c_{21}=\sqrt{\frac{p_2}{2N(1+p_1)}}$&$\text{L}_{1}$ relays in phase 2\\ \hline
$\mathbf{t}^{(2)}_{2j}=\mathbf{A}_{2j}\mathbf{r}_{2j}^{(1)}$&$c_{22}=\sqrt{\frac{p_3}{2N(1+\sigma_2^2p_1)}}$&$\text{L}_{2}$ relays in phase 2\\ \hline
$\mathbf{t}^{(3)}_{1j}=\mathbf{t}^{(2)}_{1j}$&$c_{31}=c_{21}$&$\text{L}_{1}$ relays in phase 3\\ \hline
$\mathbf{t}^{(3)}_{2j}=\mathbf{t}^{(2)}_{2j}$&$c_{32}=c_{22}$&$\text{L}_{2}$ relays in phase 3 \\ \hline 
\end{tabular}
\end{center}
\end{table}
\subsubsection{ML Decoder}
\indent The two received vectors at D for MJHS are as shown in \eqref{rdkMJHS}, which we shall call $\mathbf{x}$ for $k=2$ and $\mathbf{z}$ for $k=3$.  Let $\mathbf{y}$ be the concatenated vector of $\mathbf{x}$ and $\mathbf{z}$ namely  $\mathbf{y}=[\mathbf{x}^{\text{T}}|\mathbf{z}^{\text{T}}]^{\text{T}}$.  It can be proved as in RMC that $\mathbf{y}$ is jointly Gaussian and that the mean vector, $\mathbf{m}_y$ and covariance matrix, $\mathbf{P}_y$ of $\mathbf{y}$ are given in \eqref{meancovarianceyRMC}.  The mean, covariance, and cross-covariance of the received vectors can be proved to be
\begin{align*}
\mathbf{m}_x=&c_1\mathbf{S}_1\mathbf{h}_{s,1,d},\\
\mathbf{m}_z=&\mathbf{m}_x,\\
\mathbf{P}_x=&\left[1+c_{21}^2\sum_{j=1}^N|h_{1j,d}|^2+c_{22}^2\sum_{j=1}^N|h_{2j,d}|^2\right]\mathbf{I}_T,\\
\mathbf{P}_z=&\left[1+c_{31}^2\sum_{j=1}^N|h_{1j,d}|^2+c_{32}^2\sum_{j=1}^N|h_{2j,d}|^2\right]\mathbf{I}_T, \text{ and}\\
\mathbf{P}_{xz}=&\left[c_{21}c_{31}\sum_{j=1}^N|h_{1j,d}|^2+c_{22}c_{32}\sum_{j=1}^N|h_{2j,d}|^2\right]\mathbf{I}_T.
\end{align*}
The decoded vector is given by 
\begin{align}
\widehat{\mathbf{s}}=\arg\max_{\mathbf{s}}\Pr(\mathbf{y}|\mathbf{s})=\arg\min_{\mathbf{s}}\left\|\mathbf{y}'\right\|^2
\end{align}
where 
$\mathbf{y}'=\mathbf{P}_y^{-\frac{1}{2}}(\mathbf{y}-\mathbf{m}_y)$.  
\subsubsection{Receive SNR}
From equation \eqref{rdkMJHS} we can derive the receive SNR of this protocol at D to be
\begin{align*}
\text{snr}_{\text{MJHS}}=&\frac{Nc_1^2(c_{21}^2+c_{22}^2)\sigma_2^2}{c_{21}^2NT\sigma_2^2+c_{22}^2NT+T}.
\end{align*}
This can be simplified to 
\begin{align}
\text{snr}_{\text{MJHS}}=
\frac{p_1\sigma_2^2\left[(1 + p_1)p_3+p_2(1+p_1\sigma_2^2)\right]}{\begin{array}{c}2+ p_3+2p_1^2\sigma_2^2+p_2\sigma_2^2\\+p_1(2 +p_3+2\sigma_2^2+p_2\sigma_2^4)\end{array}}.
 \label{snrMJHS}
\end{align}
Maximizing the receive SNR shown in \eqref{snrMJHS} became quite tedious and hence  a fine computer search has been resorted to, as discussed in Section \ref{optpowallocMJHS}.  Optimum power allocation equations have been obtained by curve fitting as a function of the total average power in that Section.
\subsection{Relay SNR Combining (RSC)}\label{RSCintro}
Various phases of RSC are similar to that of RMC shown in Fig.~\ref{fig:RMCRSCRMCKCphases} and explained in the first paragraph of Section \ref{RMCintro}.  As the name suggests, in this protocol the relays in the L$_2$ layer combine the two received vectors using the respective SNRs.
\subsubsection{Protocol Analysis}
Here every operation till second phase is the same like RMC, but at Layer L$_2$ the relays R$_{2j}$ $(1 \leq j \leq N)$ combine the two vectors $\mathbf{r}_{2j}^{(1)}$ and $\mathbf{r}_{2j}^{(2)}$ in a different fashion for transmission.  
The vector that is transmitted is
\begin{align*}
c_3\mathbf{t}_{2j}^{(3)}=c_3\mathbf{A}_{2j}\left[\gamma_1\mathbf{r}_{2j}^{(1)}+\gamma_2\mathbf{r}_{2j}^{(2)}\right].
\end{align*}
Here $\gamma_1$ and $\gamma_2$ are the SNRs of the received signals $\mathbf{r}_{2j}^{(1)}$ and $\mathbf{r}_{2j}^{(2)}$ respectively at R$_{2j}$.  
These can be derived to be
\begin{align}
\gamma_1=p_1\sigma_2^2 \text{ and } \gamma_2=\frac{p_1p_2}{1+p_1+p_2}.
\end{align}
$c_1$ is the same as that shown in \eqref{c1} and $c_2$ is similar to that of RMC protocol as shown in equation \eqref{c2RMC}.  Let us work out $c_3$ with the restriction that the power transmitted by each of the L$_2$ relays is $p_3T/N$ in $T$ duration.  The power transmitted is 
\begin{align}
\frac{p_3T}{N}=&E[c_3^2\mathbf{t}_{2j}^{(3)H}\mathbf{t}_{2j}^{(3)}]\nonumber \\
=&c_3^2\left[\gamma_1^2p_1T\sigma_2^2+(\gamma_1^2+\gamma_2^2)T+\gamma_2^2p_2T\right] \nonumber \\
\Rightarrow c_3=&\sqrt{\frac{p_3}{N[\gamma_1^2(1+p_1\sigma_2^2)+\gamma_2^2(1+p_2)]}}. \label{c3RSC}
\end{align}
In phase 2 the received vectors are the same as that shown in equations \eqref{r2j2RMC} and \eqref{rd2RMC} for L$_2$ layers and D respectively.  In phase 3, it can be shown that  the destination receives
\begin{align}
\mathbf{r}_d^{(3)}=&c_1c_3\gamma_1\mathbf{S}_2\mathbf{h}_{s,2,d}  + \mathbf{u}_{z} \nonumber \\
+&c_1c_2c_3\gamma_2[\mathbf{S}_{21,1}\mathbf{h}_{s,1,21} \ldots 
\mathbf{S}_{2N,1}\mathbf{h}_{s,1,2N}]\mathbf{h}_{2,d}
\label{rd3RSC}
\end{align}
where
\begin{align*}
\mathbf{S}_2=&\left[\mathbf{A}_{21}\mathbf{s} \ldots \mathbf{A}_{2N}\mathbf{s}\right], \\ \mathbf{S}_{2n,1}=&[\mathbf{A}_{2n}\mathbf{A}_{11}\mathbf{s} \ldots \mathbf{A}_{2n}\mathbf{A}_{1N}\mathbf{s}]
\end{align*}
 and
\begin{align}
\mathbf{u}_{z}=&c_3\gamma_1\sum_{j=1}^N\mathbf{A}_{2j}\mathbf{u}_{2j}^{(1)}h_{2j,d}+c_3\gamma_2\sum_{j=1}^N\mathbf{A}_{2j}\mathbf{u}_{2j}^{(2)}h_{2j,d} \nonumber \\ 
+&c_2c_3\gamma_2\sum_{i=1}^N\sum_{j=1}^N\mathbf{A}_{2i}\mathbf{A}_{1j}\mathbf{u}_{1j}^{(1)}h_{1j,2i}h_{2i,d}+\mathbf{u}_d^{(3)} \label{uzRSC}
\end{align}
with $\mathbf{h}_{s,1,2n}$ given in \eqref{h2dRMC}.
\begin{table}[htb]
\begin{center}
\caption{Transmitted vectors and the multiplication factors - RSC} \label{tableRSC}
\vspace{0.2cm}
\begin{tabular}{|c|c|c|}\hline
\tt{Vector}&\tt{Factor}&\tt{Transmitted by}\\ \hline\hline
$\mathbf{s}$&$c_1=\sqrt{p_1T}$&S in phase 1\\ \hline
$\mathbf{t}^{(2)}_{1j}$&$c_2=\sqrt{\frac{p_2}{N(p_1+1)}}$&$\text{L}_{1}$ relays in phase 2\\ \hline
$\mathbf{t}^{(3)}_{2j}$&$c_3=\sqrt{\frac{p_3}{N\left[\gamma_1^2(1+p_1\sigma_2^2)+\gamma_2^2(1+p_2)\right]}}$&$\text{L}_{2}$ relays in phase 3\\ \hline
\end{tabular}
\end{center}
\end{table}
The transmission vectors and the corresponding factors are summarized in Table \ref{tableRSC}.
\subsubsection{ML Decoder}
\indent The two received vectors at D for RSC are as shown in \eqref{rd2RMC} and \eqref{rd3RSC} which we shall call $\mathbf{x}$ and $\mathbf{z}$ respectively.  Let $\mathbf{y}$ be the concatenation of these vectors, namely,  $\mathbf{y}=[\mathbf{x}^{\text{T}}|\mathbf{z}^{\text{T}}]^{\text{T}}$.  It can be proved as in RMC that $\mathbf{y}$ is jointly Gaussian and that the mean vector, $\mathbf{m}_y$ and covariance matrix, $\mathbf{P}_y$ of $\mathbf{y}$ are given in \eqref{meancovarianceyRMC}.  Here $\mathbf{m}_x$ and $\mathbf{P}_x$ are the same as that shown in \eqref{mxRMC} and \eqref{PxRMC} respectively.  Also the mean vector and covariance matrix of $\mathbf{z}$ along with cross covariance matrix can be proved to be
\begin{align}
\mathbf{m}_z=&c_1c_3\gamma_1\mathbf{S}_2\mathbf{h}_{s,2,d}  \nonumber \\
+&c_1c_2c_3\gamma_2\left[\mathbf{S}_{21,1}\mathbf{h}_{s,1,21} \ldots 
\mathbf{S}_{2N,1}\mathbf{h}_{s,1,2N}\right]\mathbf{h}_{2,d} \\
\mathbf{P}_z=&\left[1+c_3^2(\gamma_1^2+\gamma_2^2)\sum_{j=1}^N|h_{2j,d}|^2\right]\mathbf{I}_T\nonumber \\ +&c_2^2c_3^2\gamma_2^2\sum_{i=1}^N\sum_{j=1}^N\sum_{k=1}^N\mathbf{A}_{2i}\mathbf{A}_{2k}^{\text{T}}h_{1j,2i}h_{2i,d}h_{1j,2k}^*h_{2k,d}^*,\\
\text{and}\nonumber \\
\mathbf{P}_{xz}=&c_2^2c_3\gamma_2\sum_{j=1}^N\sum_{k=1}^Nh_{1k,d}h_{1k,2j}^*h_{2j,d}^*\mathbf{A}_{2j}^{\text{H}}=\mathbf{P}_{zx}^\text{H}.
\end{align}
The decoded vector is given by 
\begin{align}
\widehat{\mathbf{s}}=\arg\max_{\mathbf{s}}\Pr(\mathbf{y}|\mathbf{s})=\arg\min_{\mathbf{s}}\left\|\mathbf{y}'\right\|^2
\end{align}
where 
$\mathbf{y}'=\mathbf{P}_y^{-\frac{1}{2}}(\mathbf{y}-\mathbf{m}_y)$.  
\subsubsection{Receive SNR}
The receive SNR can be derived for this protocol to be
\begin{align*}
\text{snr}_{\text{RSC}}=& \frac{c_1^2c_2^2N\sigma_2^2+c_1^2c_3^2\gamma_1^2N\sigma_2^2+c_1^2c_2^2c_3^2\gamma_2^2N^2}{c_2^2NT\sigma_2^2+c_3^2(\gamma_1^2+\gamma_2^2)NT+c_2^2c_3^2\gamma_2^2TN^2+2T} \end{align*}
\begin{figure*}[!t]
\normalsize
\begin{eqnarray}
\text{snr}_{\text{RSC}}& = & \frac{\begin{array}{c}p_1p_2\sigma_2^2\left[p_2^2(1+p_2)+(\sigma_2^4+p_1\sigma_2^6)(1+p_1+p_2)^2\right]+p_1p_2^3p_3+p_1p_3\sigma_2^6(1+p_1)(1+p_1+p_2)^2\end{array}}{\begin{array}{c}2(1+p_1)\left[p_2^2(1+p_2)+(\sigma_2^4+p_1\sigma_2^6)(1+p_1+p_2)^2\right]+p_2\sigma_2^2\left[p_2^2(1+p_2)+(\sigma_2^4+p_1\sigma_2^6)(1+p_1+p_2)^2\right]+p_2^3p_3\end{array}}.\nonumber \\
\label{snrRSC}
\end{eqnarray}
\hrulefill
\vspace*{4pt}
\end{figure*}
which is simplified and shown in equation \eqref{snrRSC} at the top of next page.  Maximizing the receive SNR shown in \eqref{snrRSC} is quite tedious and hence  a fine computer search was resorted to, as discussed in Section \ref{optpowalloc}.
\subsection{RMC with Known Channel (RMCKC)}
Various phases of RMCKC are similar to that of RMC shown in Fig.~\ref{fig:RMCRSCRMCKCphases} and explained in the first paragraph of Section \ref{RMCintro}.
In this protocol the relays R$_{ij}$ are presumed to know the receive channels; R$_{1j}$ knows $h_{s,1j}$, and R$_{2j}$ knows $h_{s,2j}$ and $h_{1i,2j},~i \in \{1,\ldots,N\}$. \emph{Note that in RMCKC the relays do not know the transmit channels $h_{ij,d}$.}\\
\subsubsection{Protocol Analysis}
In phase 2, the L$_1$ relays transmit $c_2\mathbf{t}_{1j}^{(2)}$ where $\mathbf{t}_{1j}^{(2)}=\mathbf{A}_{1j}\mathbf{r}_{1j}^{(1)}h_{s,1j}^*.$  Here $c_2$ is similar to that of RMC shown in \eqref{c2RMC}.
Now L$_2$ layer relays would transmit $c_3\mathbf{t}_{2j}^{(3)}$ where $\mathbf{t}_{2j}^{(3)}=\mathbf{A}_{2j}^{'}\mathbf{r}_{2j}$ and $\mathbf{r}_{2j}$ is a concatenated vector given by
\begin{align*}
\mathbf{r}_{2j}=
\begin{bmatrix}
\mathbf{r}_{2j}^{(1)}h_{s,2j}^* \\
\mathbf{r}_{2j}^{(2)}\left\|\mathbf{h}_{1,2j}\right\| \\
\end{bmatrix}.
\end{align*}
Also the received vector at R$_{2j}$ in phase 2 is
\begin{align}
\mathbf{r}_{2j}^{(2)}=&\sum_{i=1}^Nc_2\mathbf{t}_{1i}^{(2)}h_{1i,2j}+\mathbf{u}_{2j}^{(2)}\nonumber \\
=&c_1c_2\sum_{i=1}^N|h_{s,1i}|^2h_{1i,2j}\mathbf{A}_{1i}\mathbf{s}\nonumber \\
+&c_2\sum_{i=1}^Nh_{s,1i}^*h_{1i,2j}\mathbf{A}_{1i}\mathbf{u}_{1i}^{(1)}+\mathbf{u}_{2j}^{(2)}.
\end{align}
Here $\mathbf{A}_{2j}'$ is the same as that of RMC shown in \eqref{A2jprimeRMC}.  The multiplying factor $h_{s,2j}^*$ is the conjugate of the channel the transmitted signal would have gone through when $\mathbf{r}_{2j}^{(1)}$ is received.  Similarly, the transmitted signal would have gone through a vector of channel coefficients $\mathbf{h}_{1,2j}$ when $\mathbf{r}_{2j}^{(2)}$ is received, and hence $\left\|\mathbf{h}_{1,2j}\right\|$ is the multiplying factor. \\
\indent The received vector $\mathbf{r}_{d}^{(2)}$ at D can be proved to be
\begin{align}
\mathbf{r}_{d}^{(2)}=c_1c_2[\mathbf{A}_{11}\mathbf{s} \ldots \mathbf{A}_{1N}\mathbf{s}]\mathbf{h}_{s,1,d}'+\mathbf{u}_{x} \label{rd2RMCKC}
\end{align} 
where 
\begin{align}
\mathbf{h}_{s,1,d}'=&
\begin{bmatrix}
\left|h_{s,11}\right|^2h_{11,d} \\
\vdots \\
\left|h_{s,1N}\right|^2h_{1N,d}
\end{bmatrix}\text{ and } \nonumber \\
\mathbf{u}_{x}=&c_2\sum_{j=1}^Nh_{s,1j}^*h_{1j,d}\mathbf{A}_{1j}\mathbf{u}_{1j}^{(1)}+\mathbf{u}_d^{(2)}. \label{uxRMCKC}
\end{align}
The received vector $\mathbf{r}_{d}^{(3)}$ at D can be proved to be 
\begin{align}
\mathbf{r}_d^{(3)}=&\frac{c_1c_3}{\sqrt{2}}\sum_{j=1}^N\left|h_{s,2j}\right|^2h_{2j,d}\mathbf{A}_{2j}(1)\mathbf{s} \nonumber \\ +&\frac{c_1c_2c_3}{\sqrt{2}}\sum_{i=1}^N\sum_{j=1}^N\left\|\mathbf{h}_{1,2j}\right\|h_{2j,d}h_{1i,2j}\left|h_{s,1i}\right|^2 \mathbf{A}_{2j}(2)\mathbf{A}_{1i}\mathbf{s}\nonumber \\
+& \mathbf{u}_{z} \label{rd3RMCKC} \\
\text{ where} \nonumber \\
\mathbf{u}_{z}=&\frac{c_3}{\sqrt{2}}\sum_{j=1}^Nh_{s,2j}^*h_{2j,d}\mathbf{A}_{2j}(1)\mathbf{u}_{2j}^{(1)} \nonumber \\ +&\frac{c_2c_3}{\sqrt{2}}\sum_{j=1}^N\sum_{i=1}^N\left\|\mathbf{h}_{1,2j}\right\|h_{2j,d}h_{s,1i}^*h_{1i,2j}\mathbf{A}_{2j}(2)\mathbf{A}_{1i}\mathbf{u}_{1i}^{(1)} \nonumber \\
+&\frac{c_3}{\sqrt{2}}\sum_{j=1}^N\left\|\mathbf{h}_{1,2j}\right\|h_{2j,d}\mathbf{A}_{2j}(2)\mathbf{u}_{2j}^{(2)}+\mathbf{u}_d^{(3)}.
\end{align}
With the total average power transmitted per symbol duration fixed at $p_3/N$ in phase 3, $c_3$ can be derived to be
\begin{eqnarray}
c_3 = \sqrt{\frac{(1+p_1)p_3}{\begin{array}{c} N[8p_1p_2 + N(1+p_1+p_2)\\+(1+p_1)\sigma_2^2 +\sigma_2^4p_1 (1+p_1)]\end{array}}}\label{c3RMCKC}
\end{eqnarray}
Expressions for $c_1, c_2, c_3$, and the transmission vectors are summarized in Table \ref{tableRMCKC}.  
\begin{table}[htb]
\begin{center}
\caption{Transmitted vectors and multiplication factors - RMCKC} \label{tableRMCKC}
\vspace{0.2cm}
\begin{tabular}{|c|c|c|}\hline
\tt{Vector}&\tt{Factor}&\tt{Transmitted by}\\ \hline\hline
$\mathbf{s}$&$c_1=\sqrt{p_1T}$&S in phase 1\\ \hline
$\mathbf{t}^{(2)}_{1j}$&$c_2=\sqrt{\frac{p_2}{N(p_1+1)}}$&$\text{L}_{1}$ relays in phase 2\\ \hline
$\mathbf{t}^{(3)}_{2j}$&$c_3$, shown in \eqref{c3RMCKC}&$\text{L}_{2}$ relays in phase 3\\ \hline
\end{tabular}
\end{center}
\end{table}
\subsubsection{ML Decoder}
\indent The two received vectors at D for RMCKC are as shown in \eqref{rd2RMCKC} and \eqref{rd3RMCKC}, which we shall call $\mathbf{x}$ and $\mathbf{z}$ respectively.  Let $\mathbf{y}$ be the concatenated vector of $\mathbf{x}$ and $\mathbf{z}$ namely  $\mathbf{y}=[\mathbf{x}^{\text{T}}|\mathbf{z}^{\text{T}}]^{\text{T}}$.  It can be proved as in RMC that $\mathbf{y}$ is jointly Gaussian and that the mean vector, $\mathbf{m}_y$ and covariance matrix, $\mathbf{P}_y$ of $\mathbf{y}$ are given in \eqref{meancovarianceyRMC}.  The mean vector, covariance, and cross covariance matrices can be proved to be
\begin{align}
\mathbf{m}_x=&c_1c_2\sum_{i=1}^N|h_{s,1i}|^2h_{1i,d}\mathbf{A}_{1i}\mathbf{s},\\
\mathbf{P}_x=&\left[1+c_2^2\sum_{i=1}^N|h_{s,1i}|^2|h_{1i,d}|^2\right]\mathbf{I}_T,\\
\mathbf{m}_z=&\frac{c_1c_3}{\sqrt{2}}\sum_{j=1}^N\left|h_{s,2j}\right|^2h_{2j,d}\mathbf{A}_{2j}(1)\mathbf{s}\nonumber \\ +&\frac{c_1c_2c_3}{\sqrt{2}}\sum_{i=1}^N\sum_{j=1}^N\left\|\mathbf{h}_{1,2j}\right\|h_{2j,d}h_{1i,2j}\left|h_{s,1i}\right|^2 \mathbf{A}_{2j}(2)\mathbf{A}_{1i}\mathbf{s}, \\
\mathbf{P}_z=&
\left[1+\frac{c_3^2}{2}\sum_{j=1}^N\left(|h_{s,2j}|^2
+\left\|\mathbf{h}_{1,2j}\right\|^2\right)|h_{2j,d}|^2\right]\mathbf{I}_T\nonumber \\
+&\frac{c_2^2c_3^2}{2}\sum_{i=1}^N\sum_{j=1}^N\sum_{k=1}^N \left\|\mathbf{h}_{1,2j}\right\|\left\|\mathbf{h}_{1,2k}\right\| h_{2j,d}h_{2k,d}^*|h_{s,1i}|^2\nonumber \\
&h_{1i,2j}h_{1i,2k}^*\mathbf{A}_{2j}(2)\mathbf{A}_{2k}^\text{T}(2),  \text{ and} \\
\mathbf{P}_{xz}=&\frac{c_2^2c_3}{\sqrt{2}}\sum_{j=1}^N\sum_{i=1}^N\left\|\mathbf{h}_{1,2j}\right\|h_{2j,d}^*|h_{s,1i}|^2h_{1i,d}h_{1i,2j}^*\mathbf{A}_{2j}^{\text{T}}(2).
\end{align}
The decoded vector is given by 
\begin{align}
\widehat{\mathbf{s}}=\arg\max_{\mathbf{s}}\Pr(\mathbf{y}|\mathbf{s})=\arg\min_{\mathbf{s}}\left\|\mathbf{y}'\right\|^2
\end{align}
where 
$\mathbf{y}'=\mathbf{P}_y^{-\frac{1}{2}}(\mathbf{y}-\mathbf{m}_y)$.  
\subsubsection{Receive SNR}
From equations \eqref{rd2RMCKC} and \eqref{rd3RMCKC}, we can derive the receive SNR of this protocol at D to be
\begin{align*}
\text{snr}_{\text{RMCKC}}=&\frac{16Nc_1^2 c_2^2\sigma_2^2+2Nc_1^2c_3^2(3\sigma_2^4+\sigma_2^2)+8N^2c_1^2c_2^2c_3^2}{2NTc_2^2\sigma_2^2+4T+N^3c_2^2c_3^2T+N^2c_3^2T+Nc_3^2\sigma_2^2T}.
\end{align*}
\begin{figure*}[!t]
\normalsize
\begin{eqnarray}+
\text{snr}_{\text{RMCKC}}& \nonumber \\
=&\frac{\begin{array}{c}16(N+8p_1)p_1p_2^2\sigma_2^2 +2p_1p_3\sigma_2^2(1+p_1)^2(1+3\sigma_2^2)+8p_1p_2(1+p_1)[p_3+2\sigma_2^2(N+\sigma_2^2+p_1\sigma_2^4)]\end{array}}{\begin{array}{c}(1+p_1)\left[N(1+p_2)(4+p_3)+p_1(32p_2+N(4+p_3))\right]+ 4(1+p_1)^2\sigma_2^2+2N(1+p_1)p_2\sigma_2^2+\\2(N+8p_1)p_2^2\sigma_2^2+(1+p_1)^2p_3\sigma_2^2 +2(1+p_1)\left[2p_1(1+p_1)+p_2\right]\sigma_2^4+2p_1(1+p_1)p_2\sigma_2^6
\end{array}}.\label{snrRMCKC}
\end{eqnarray}
\hrulefill
\vspace*{4pt}
\end{figure*}
$\text{snr}_{\text{RMCKC}}$ is simplified and shown in equation \eqref{snrRMCKC} at the top of next page. Maximizing the receive SNR shown in \eqref{snrRMCKC} became quite tedious and hence a fine computer search has been resorted to, as discussed in Section \ref{optpowalloc}.

%% file: sec_simulations.tex
\section{Simulations} \label{simulations}
We have seen five different protocols, namely RMC, EJHS, MJHS, RSC, and RMCKC.  In all these protocols matrices at relays have been used, for generating a distributed space-time code.  Using simulations the performance of the system has been compared, when these matrices are real orthogonal and complex unitary.  Optimum power allocation to all transmissions using simulations have been found out.  Finally BERs for various protocols have been plotted while using the optimum power allocations obtained.\\
\indent In the simulations, a block size of length $T=5$ symbol duration and number of relays in each layer, $N=5$ for a run of 10,000 data blocks have been used.  As defined earlier $\mathbf{s}=[s(1)\cdots s(T)]^{\text{T}},$ and $s(k)=s_r(k)+js_i(k),~1 \leq k \leq T$. Let us also assume that the real part $s_r(k)$ and the imaginary part $s_i(k)$ of $s(k)$ are equally likely selected from the $M$-PAM signal set 
\begin{align*}
K\left \{-\frac{M-1}{2},\cdots,-\frac{1}{2},\frac{1}{2},\cdots,\frac{M-1}{2}\right\},  
\end{align*}
where $K$ is the normalizing factor so that $E\left[\mathbf{s}^{\text{H}}\mathbf{s}\right]=1.$  Hence the cardinality, $L$, of $\Omega$ is $M^{2T}$.  The value of $K$ is found from
\begin{align*}
E\left[\mathbf{s}^{\text{H}}\mathbf{s}\right]=&E\sum_{k=1}^T|s(k)|^2=TE\left[|s(k)|^2\right]\\=&TE\left[s_r^2(k)+s_i^2(k)\right]\\
=&2TE\left[s_r^2(k)\right]\\
=&\frac{TK^2}{M}\sum_{j=1}^{M/2}(2j-1)^2=\frac{TK^2}{M}\frac{(M-1)M(M+1)}{6}\\
=&1 \Rightarrow K=\sqrt{\frac{6}{T(M^2-1)}}.
\end{align*}
$M=2$ has been used in all the simulations.
\subsection{Relay Matrices}
The relay matrices $\mathbf{A}_{ij}$ have been selected to be real orthogonal as the performance in terms of BER is the same as that when complex unitary matrices are used \cite{jing}. 
\begin{figure}[!tb]
\centering
\includegraphics[width=0.4 \textwidth]{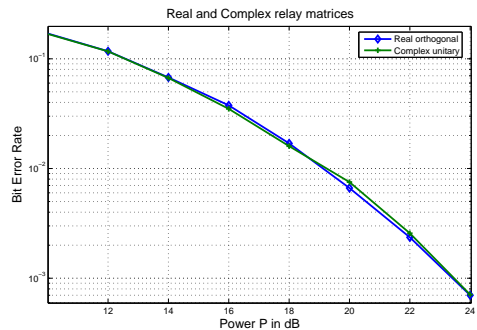}
\caption{Comparison of performance of real orthogonal matrices with complex unitary matrices.}\label{fig:BER10000T5N5ComplexRealA10to24}
\end{figure}
To prove this, simulations were carried out with the simple JHS system. Fig.~\ref{fig:BER10000T5N5ComplexRealA10to24} shows a plot of transmitted power vs. BER achieved where there are two curves one representing that of using real orthogonal and the other complex unitary matrices at the relays.  It is clear that the BER for all SNRs using real is the same as that while using complex matrices.  Hence in all the simulations, real orthogonal matrices have been used to make DSTC.
\subsection{Optimum Power Allocation} \label{optpowalloc}
\begin{figure}[!tb]
\centering
\includegraphics[width=0.4 \textwidth]{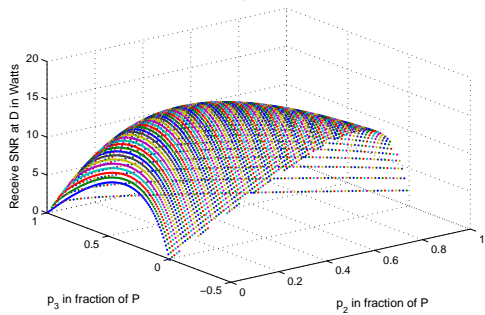}
\caption{Plot of receive SNR for RMC.}\label{fig:plotreceivesnrRMC}
\end{figure}

\begin{figure}[!tb]
\centering
\includegraphics[width=0.4 \textwidth]{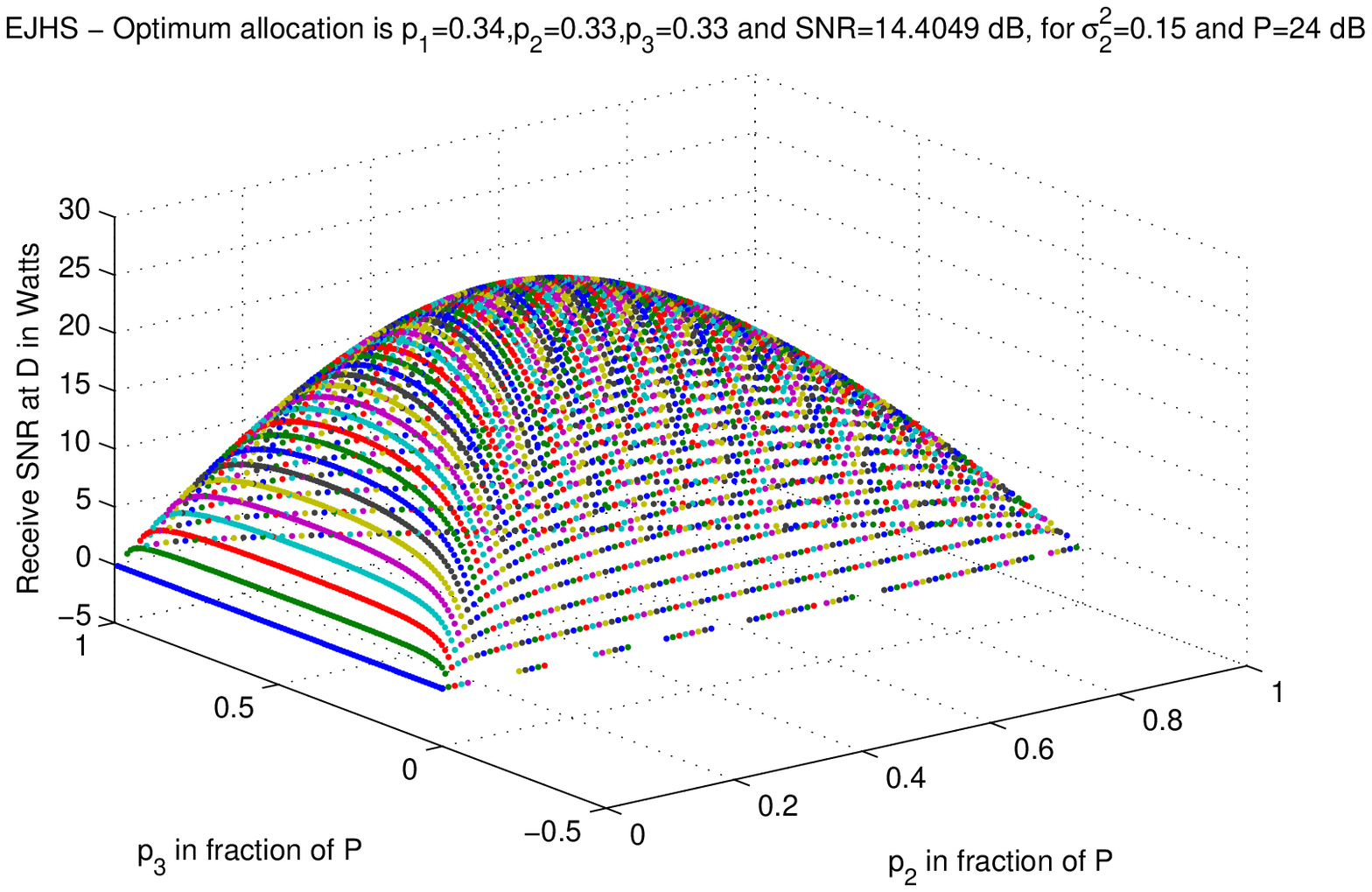}
\caption{Plot of receive SNR for EJHS.}\label{fig:plotreceivesnrEJHS}
\end{figure}

\begin{figure}[!tb]
\centering
\includegraphics[width=0.4 \textwidth]{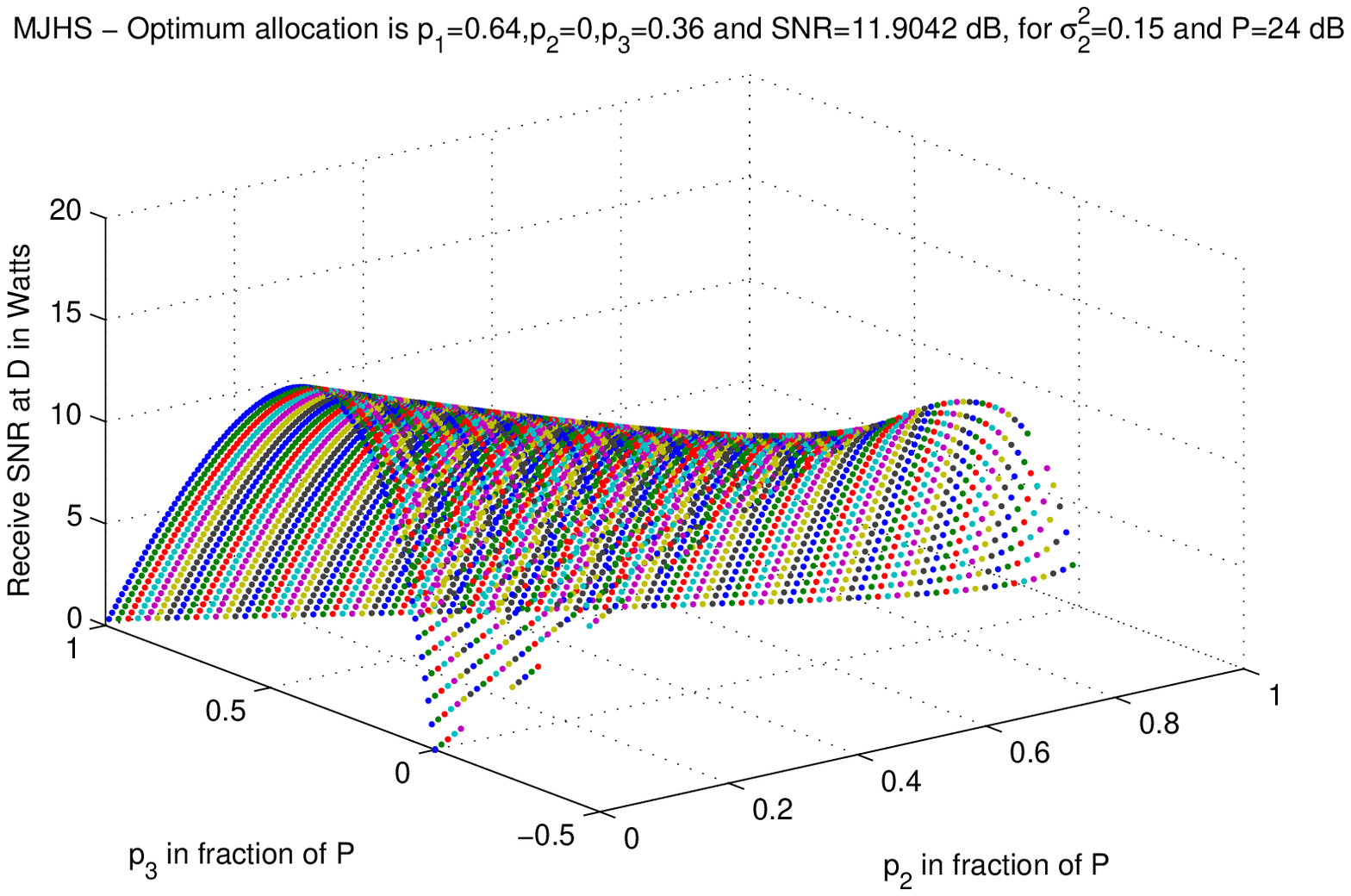}
\caption{Plot of receive SNR for MJHS.}\label{fig:plotreceivesnrMJHS}
\end{figure}

\begin{figure}[!tb]
\centering
\includegraphics[width=0.4 \textwidth]{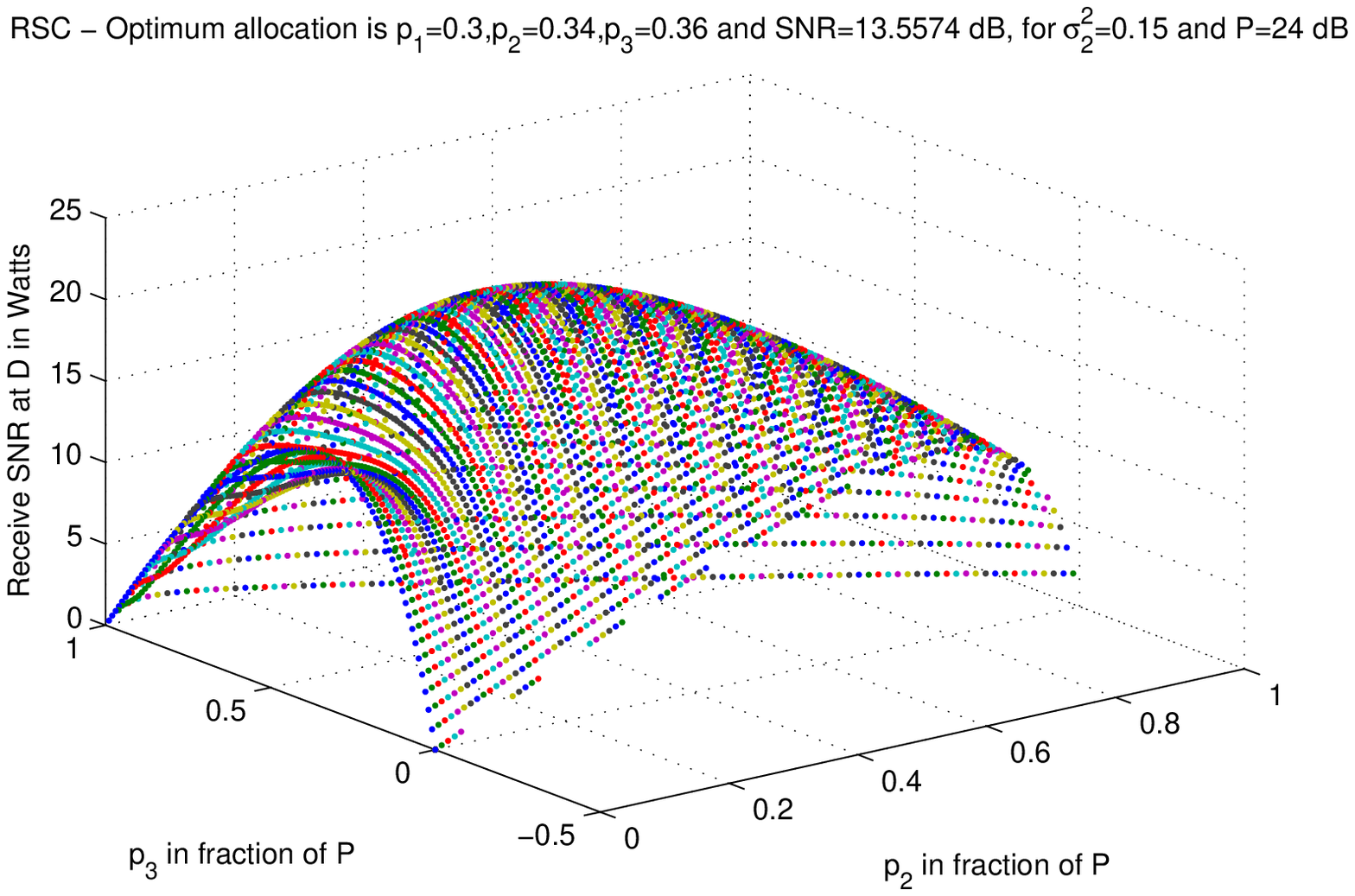}
\caption{Plot of receive SNR for RSC.}\label{fig:plotreceivesnrRSC}
\end{figure}

\begin{figure}[!tb]
\centering
\includegraphics[width=0.4 \textwidth]{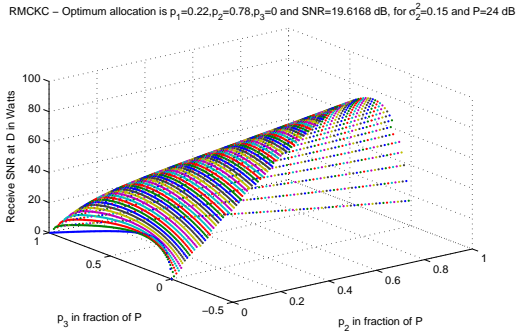}
\caption{Plot of receive SNR for RMCKC.}\label{fig:plotreceivesnrRMCKC}
\end{figure}

\begin{figure}[!tb]
\centering
\includegraphics[width=0.4 \textwidth]{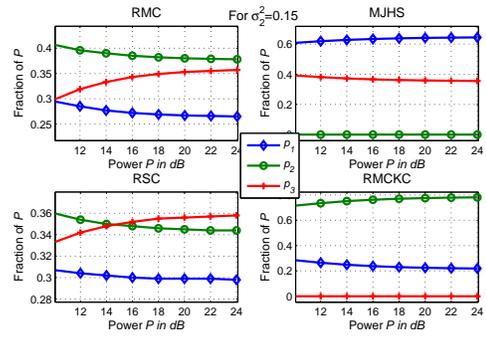}
\caption{Plot of Optimum power allocations for MJHS, RMC, RSC, and RMCKC for $\sigma_2^2=0.15$.}\label{Optpowalloc4protocolsVar0_15}
\end{figure}

\begin{figure}[!tb]
\centering
\includegraphics[width=0.4 \textwidth]{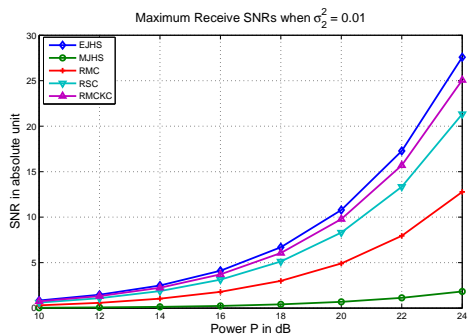}
\caption{Plot of Maximum Receive SNRs for  $\sigma_2^2=0.01$.}\label{fig:FiveprotocolsSNRVar0_01}
\end{figure}

Allocation of power to various transmissions, namely, $p_1, p_2$, and $p_3$ are to be done in such a way that it minimizes the transmission errors.  Ideally one should  minimize probability of error (PE) or pairwise error probability (PEP) and obtain the optimum power allocation.   Computation of PEP was found to be complicated.   In \cite{jing} the authors proved that the optimum power allocation obtained by minimizing PEP also maximizes receive SNR for their system model and protocol. Hence in this work, receive SNR has been selected as the parameter to be maximized and expect that this gives near optimum power allocation.  Maximizing this receive SNR analytically became too complex and hence a fine computer search has been carried out as explained here.\\
\indent We have 3 variables namely $p_1, p_2,$ and $p_3$ which are the powers allocated to the three transmissions used in the protocols discussed.  These three variables have two constraints,  namely, $p_1+p_2+p_3 \leq P$ and $p_1, p_2, p_3 \geq 0$.  Let us consider the best case constraint of $p_1+p_2+p_3 = P$.  
\begin{figure}[!tb]
\centering
\includegraphics[width=0.4 \textwidth]{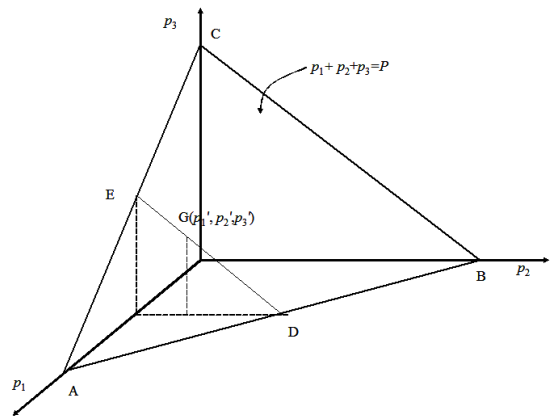}
\caption{Power distribution surface where optimum power allocation point resides.}\label{fig:Optpowalloc}
\end{figure}
This can be geometrically expressed as shown in Fig.~\ref{fig:Optpowalloc}, where AB is in $p_1-p_2,$ BC in $p_2-p_3$ and AC in $p_1-p_3$ planes.  We can select $p_1$ and keep varying $p_2$ with $p_3$ automatically getting fixed.  All the points on this plane need to be considered to find the optimum power allocation.  As it is impossible to consider all the points on this plane we can select them with a granularity.  Consider the straight line shown on the plane in Fig.~\ref{fig:Optpowalloc}, DE, which is parallel to BC.  The equation of this straight line is $p_1+p_2+p_3 = P$; $p_1=p_1'$ where $0 \leq p_1' \leq P.$  By varying $p_1'$, we will get more straight lines parallel to BC.  With a certain granularity we will vary $p_1'$.  i.e. $p_1' = n \delta P$ where  $0 \leq \delta \leq 1$ and $0 \leq n \leq \left\lfloor\frac{1}{\delta}\right\rfloor,~n$ being an integer.  Once  $p_1'$ is selected, let us select $p_2$ with a granularity as for the case of $p_1$ as $p_2'=m \epsilon P$ with $0 \leq m \leq \left\lfloor \frac{1-\delta}{\epsilon}\right\rfloor, \; m$ being an integer and $0 \leq \epsilon \leq 1.$ Then $p_3$ is fixed as $p_3'=P-p_1'-p_2'$.   Hence we can get the point G$(p_1',p_2',p_3')$ as shown in Fig.~\ref{fig:Optpowalloc}.
The complete region of the plane ABC is  scanned fully and receive SNRs are calculated for each point.  That point which has the maximum SNR is selected as the optimum point.\\
\indent In the calculations, $\delta=1/1000$ and $\epsilon=1/1000$ have been used.  Hence the region has been scanned with a granularity of 0.001 in all the three axes.
The optimum points $(p_1^{opt}, p_2^{opt}, p_3^{opt})$ differed for various powers and $\sigma_2^2$.  \emph{In all the protocols $p_1, p_2,$ and $p_3$ represent the powers allocated to the three phases except in MJHS, where $p_2$ represents the power transmitted by L$_1$ relays in both second and third phases and $p_3$ represents that of L$_2$ relays in both phases}. 
\indent 3D plots of receive SNR for all the protocols for various total average power $P$, when $\sigma_2^2=0.01,~ 0.1,$ and 0.5 have been generated and obtained the optimum points when the receive SNR is maximum.  These plots for RMC, EJHS, MJHS, RSC, and RMCKC are shown in Figures \ref{fig:plotreceivesnrRMC} to \ref{fig:plotreceivesnrRMCKC} respectively for $\sigma_2^2=0.1$ and $P=24$ dB.  
The plots show receive SNR for various possible combinations of $p_1, p_2,$ and $p_3$.  It can be seen that the maximum SNR is achieved at $p_1^{opt}=0.254,~ p_2^{opt}=0.353,$ and $p_3^{opt}=0.393$ for RMC.  Also for EJHS the fact that $p_1^{opt}=1/3,~ p_2^{opt}=1/3,$ and $p_3^{opt}=1/3$ seen in subsection \ref{receiveSNREJHS} is verified from the 3D plot shown in Fig.~\ref{fig:plotreceivesnrEJHS}. 
\begin{figure}[!tb]
\centering
\includegraphics[width=0.4 \textwidth]{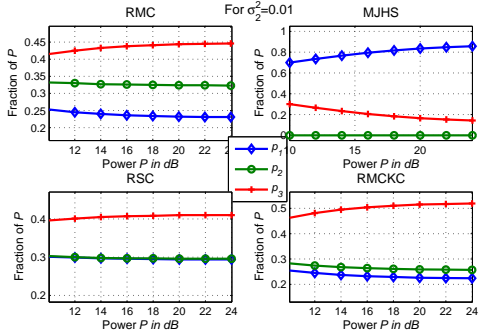}
\caption{Plot of optimum power allocations for RMC, MJHS, RSC, and RMCKC for $\sigma_2^2=0.01$.}\label{Optpowalloc4protocolsVar0_01}
\end{figure}
Figures \ref{Optpowalloc4protocolsVar0_01} to \ref{Optpowalloc4protocolsVar0_5} show plots of $p_1, p_2,$ and $p_3$ of four of the protocols RMC, MJHS, RSC, and RMCKC to achieve maximum receive SNR at D for $\sigma_2^2=0.01$, 0.1 and 0.5 respectively. (Plot for EJHS is left out as the power allocation remains the same as shown in subsection \ref{receiveSNREJHS}, for any $P$.)
The following can be observed from Fig.~\ref{Optpowalloc4protocolsVar0_01}:
\begin{itemize}
\item {RMC, RSC \& RMCKC}
\begin{itemize}
\item {$p_3$, i.e. power transmitted by L$_2$ relays, needs to be increased with the increase in $P$, whereas that of source, $p_1$, and L$_1$ relays, $p_2$ are to be reduced.}
\item{As $P$ increases the rate at which $p_1$ and $p_2$ are to be reduced or $p_3$ to be increased is less in the case of RSC compared to that of RMCKC and RMC.}
\item{Power transmitted by source and L$_1$ relays are almost the same in the case of RSC.}
\end{itemize}
\item{MJHS} \label{optpowallocMJHS}
\begin{itemize}
\item{It does not transmit using L$_1$ relays to achieve high receive SNR.  i.e. $p_2$ remains zero.}
\item{The source needs to increase its power whereas L$_2$ relays are to decrease their powers for increase in the total power.}
\item{The plots can be curve fitted by minimizing mean squared error with quadratic setting as
\begin{align} 
\widetilde{p_1}=&0.71+0.002P-4.7 \times 10^{-6}P^2, \label{p1tilde}
\end{align}
and 
\begin{align} 
\widetilde{p_3}=&0.39-0.002P+4.7 \times 10^{-6}P^2. \label{p3tilde}
\end{align}
}
\end{itemize}
\end{itemize}
The following can be observed from Fig. \ref{Optpowalloc4protocolsVar0_15}:
\begin{itemize}
\item {RMC}
\begin{itemize}
\item {The powers transmitted by source and L$_1$ layers are to be reduced while that of L$_2$ layers is to be increased as P increases.}
\item{At $P\approx10$ dB the powers transmitted by S and L$_2$ layers are the same, above which L$_2$ layer transmits more power than S.}
\end{itemize}
\item {MJHS}
\begin{itemize}
\item{To get maximum receive SNR, this protocol keeps the L$_1$ relays mute throughout.  i.e. this protocol does not require those relays that are nearer to the source.  The power transmitted by the source, $p_1$ is to be increased while that of L$_2$ relays, $p_3$ is to be decreased as the total power, $P$ increases.}
\end{itemize}
\item {RSC}
\begin{itemize}
\item{Source and L$_1$ relays are to decrease their powers while L$_2$ relays are to increase their powers to obtain maximum receive SNR, as the total power, $P$ increases.}
\item{At $P\approx15$ dB the powers transmitted by L$_1$ and L$_2$ layers are the same, above which L$_2$ layer transmits more power than S.}
\end{itemize}
\item {RMCKC}
\begin{itemize}
\item{This protocol, unlike MJHS, does not require L$_2$ relays throughout.  i.e. it keeps $p_3$ to be zero throughout.  The power of the source, $p_1$, is to be decreased while that of L$_1$ layers, $p_2$, is to be increased as the total power, $P$,  increases.}
\end{itemize}
\end{itemize}
The following can be observed from Fig. \ref{Optpowalloc4protocolsVar0_5}:
\begin{itemize}
\item {RMC/RMCKC}
\begin{itemize}
\item{Like in the case of RMCKC for $\sigma_2^2=0.1$, these protocols, for $\sigma_2^2=0.5$, do not require L$_2$ relays throughout.  i.e. $p_3$ remains to be zero throughout.  The source power, $p_1$, is to be decreased while that of L$_1$ layers, $p_2$, is to be increased as the total power, $P$, increases.}
\end{itemize}
\begin{figure}[!tb]
\centering
\includegraphics[width=0.4 \textwidth]{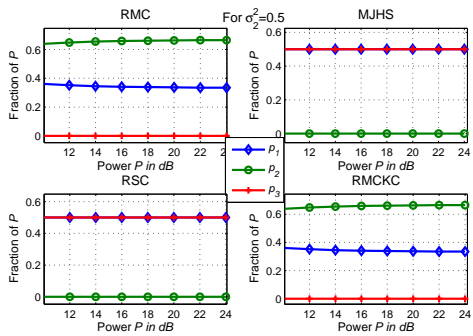}
\caption{Plot of Optimum power allocations for RMC, MJHS,  RSC, and RMCKC for $\sigma_2^2=0.5$.}\label{Optpowalloc4protocolsVar0_5}
\end{figure}
\item {MJHS/RSC}
\begin{itemize}
\item{Unlike RMC and RMCKC, these protocols do not require L$_1$ relays throughout. i.e. $p_2$ remains to be zero throughout.  Source and L$_2$ relays are allocated half the total power each, to get maximum receive SNR.}
\end{itemize}
\end{itemize}
We can infer the following from all the above observations made on figures \ref{Optpowalloc4protocolsVar0_01} to  \ref{Optpowalloc4protocolsVar0_5}:
\begin{itemize}
\item As the channels from source to L$_2$ layer and L$_1$ layer to destination improve, (i.e. $\sigma_2^2>0.01$) RMC reduces the importance to the relays in L$_2$ while giving more weightage to source and L$_1$ layer relays.
\item Irrespective of the power loss condition (i.e. for any value of $\sigma_2^2$), MJHS keeps the L$_1$ layer relays muted and does not use them throughout.  Also the difference between the powers divided between the source and L$_2$ layer relays narrows down and finally becomes zero as the channel variance from source to L$_2$ layer along with L$_1$ layer to destination improves and reaches 0.5.
\item Unlike MJHS, which shuts down L$_1$ layer relays completely in any power loss condition, RMCKC mutes L$_2$ layer relays when the signal from the source to the second layer or the L$_1$ layer relays to destination undergoes lower attenuation.
\end{itemize}
All the plots shown in figures \ref{Optpowalloc4protocolsVar0_01} to  \ref{Optpowalloc4protocolsVar0_5} can be curve fitted as shown in \eqref{p1tilde} and \eqref{p3tilde}, so that they can be readily used for power allocations.\\
\indent Fig. \ref{fig:FiveprotocolsSNRVar0_01} shows the maximum receive SNRs of the protocols
 discussed in this paper for various values of $P$ with $\sigma_2^2=0.01.$  It can be observed that the performance in terms of receive SNR of the protocols almost matches the performance in BER shown in Fig. \ref{fig:Fiveprotocols10000Var0_01} in the next Section.  
\subsection{BER Plots} \label{BERplots}
\begin{figure}[!tb]
\centering
\includegraphics[width=0.4 \textwidth]{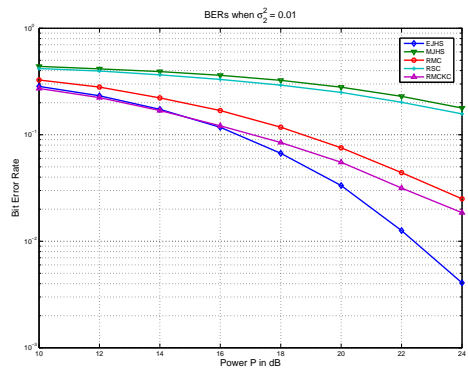}
\caption{Comparison of BER of protocols when $\sigma_2^2 = 0.01.$} \label{fig:Fiveprotocols10000Var0_01}
\end{figure}

\begin{figure}[!tb]
\centering
\includegraphics[width=0.4 \textwidth]{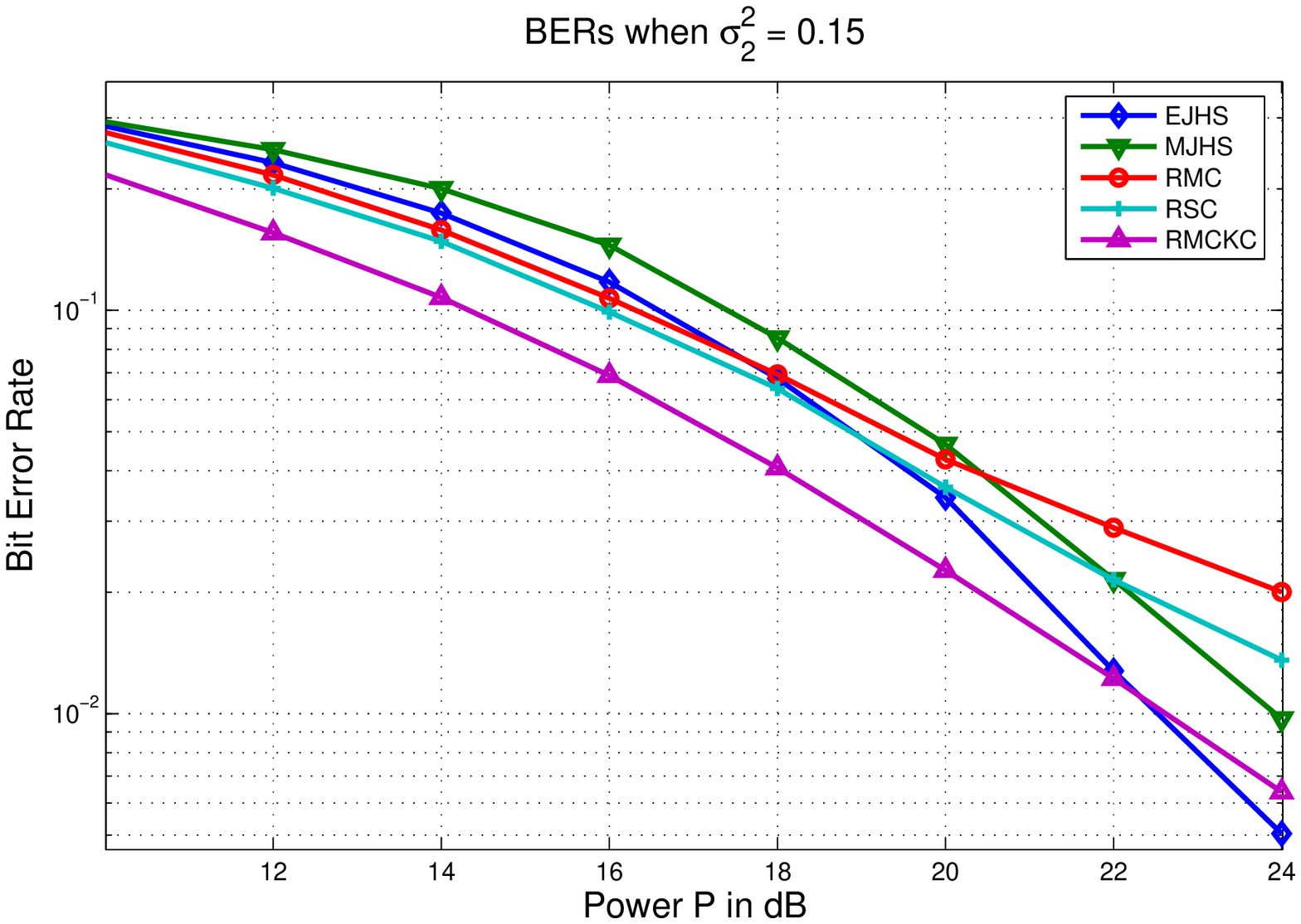}
\caption{Comparison of BER of protocols when $\sigma_2^2=0.15.$} \label{fig:Fiveprotocols10000Var0_1}
\end{figure}

\begin{figure}[!tb]
\centering
\includegraphics[width=0.4 \textwidth]{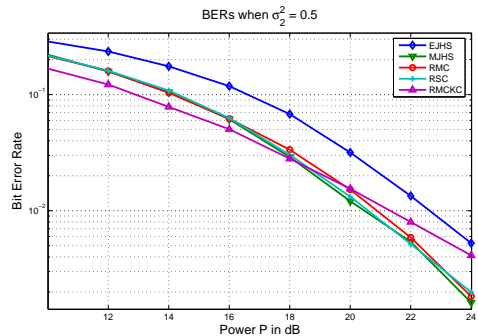}
\caption{Comparison of BER of protocols when $\sigma_2^2= 0.5.$} \label{fig:Fiveprotocols10000Var0_5}
\end{figure}

Finally for comparison of various protocols, BER plots shown in Figures \ref{fig:Fiveprotocols10000Var0_01} to \ref{fig:Fiveprotocols10000Var0_5} for $\sigma_2^2=0.01,$ 0.15, and 0.5 respectively, have been used.  These plots have been generated for all the protocols with powers allocated to each of the transmissions according to the optimum power allocation points obtained in Section \ref{optpowalloc}.
The following can be observed from Fig.~\ref{fig:Fiveprotocols10000Var0_01}:
\begin{itemize}
\item {The performance of RMCKC for $P \leq 15$ dB is the same as that of EJHS.  For $P > 15$ dB EJHS is the best.}
\item{Amongst RMC, MJHS, and RSC, RMC performs better.}
\end{itemize}
From Fig.~\ref{fig:Fiveprotocols10000Var0_1} we can observe the following:
\begin{itemize}
\item {All the protocols proposed by us except MJHS perform better than EJHS for $P \leq 18$ dB when $\sigma_2^2=0.15$.}
\item{Further the performance of RMCKC is the best for $P \leq 22$ dB.}
\end{itemize}
From Fig.~\ref{fig:Fiveprotocols10000Var0_5} we observe the following:
\begin{itemize}
\item{All the proposed protocols perform better than EJHS for $\sigma_2^2 = 0.5$ throughout the usable range of transmitted power $P$.}
\item {As expected, RMCKC attains the lowest BER using the receive channel knowledge.}
\item {For $P \geq 19$ dB, RSC, RMC, and MJHS work better than RMCKC.}
\end{itemize}
\subsection{Discussion and Observations}
\indent All the protocols proposed by us except MJHS display better performance than EJHS when $\sigma_2^2 > 0.15$ for usable range of transmitted power $P$.  Only for $\sigma_2^2 \leq 0.01$, EJHS works better than all the proposed protocols.  The reason for this is that when $\sigma_2^2$ reduces to a low value, say 0.01, the signals which reach L$_2$ in phase 1 and D in phase 2 are highly attenuated. Hence the proposed protocols, which use these attenuated signals, do not perform as good as EJHS, as some power is expended with no particular advantage in these signals.  It is also observed that for $P \geq 19$ dB, RSC, RMC, and MJHS outperform RMCKC implying that RMCKC does not use whatever channel information it has to its best and there is a possible scope for improvement.  However with just the receive channel knowledge of $h_{s,1j}$, RMCKC performs best; it has lowest BER and high data rate.\\
\indent Further when $\sigma_2^2 \rightarrow 1$ expectedly RMC and RSC have been found to perform similar to that of MJHS in which all the $2N$ relays are merged into one layer.  An interesting result which is to be emphasized is that when the signal from source to the second layer reaches with less attenuation (channel variance, $\sigma_2^2 > 0.15$), then we can opt for RMCKC which selects only those relays that are closer to the source and does not use those that are closer to the destination.  This implies that we need only two phases of transmission leading to higher data rate compared to those which use three phases.\\
\indent To summarize, when $\sigma_2^2 \leq 0.01$ we can select EJHS and no considerable gain would be obtained in going for the schemes which use `weak' links.  However, when $\sigma_2^2 > 0.01$ the proposed protocols perform better for usable range of transmitted power $P$.  Here we can select either RMC or RSC when there is no channel knowledge at the relays for lower values of $P$ depending upon $\sigma_2^2$ (e.g. for $P<18$ dB when $\sigma_2^2=0.15$).  But if the relays have just the receive channel knowledge, we can use RMCKC for most values of $P$ (e.g. for $P<19$ dB when $\sigma_2^2=0.5$, above which RMC/RSC to be used).  Also it is beneficial to select RMCKC as it gets better reliability with increased data rate, as it uses only two phases.

%% file: sec_conclusion.tex
\section{Conclusion} \label{conclusion}
\begin{figure}[h]
\centering
\includegraphics[width=0.4 \textwidth]{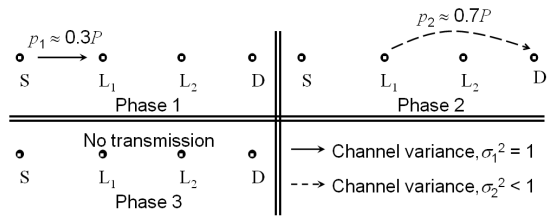}
\caption{Best scheme when $\sigma_2^2>0.15$ with receive channel knowledge at relays.} \label{resultRMCKC}
\end{figure}

In this paper, the simple relay processing system using matrices suggested by Jing and Hassibi in \cite{jing} to achieve benefits of DSTC has been modified and enlarged.  Also random orthogonal matrices have been used at relays and we have shown that BER performance achieved is the same as that when complex unitary matrices as suggested in \cite {jing} are used.  \\
\indent Four new protocols have been derived from the one proposed in \cite{jing}.  We have made use of the signals from `weak' channels (which are received by relays and destination with high power loss) in these protocols and shown that they perform better than the basic protocol proposed in \cite{jing} with reasonable strength of the `weak' channels.\\
\indent An interesting result when the relays have the receive channel knowledge in the protocol RMCKC is shown in Fig. \ref{resultRMCKC}.  Above $\sigma_2^2 \geq 0.15$, RMCKC uses only two phases.  Hence the data rate is improved by 1/3 compared to all other protocols and it gets the lowest BER also for most of the usable range of the transmitted power.

%% file: dstc_arxiv.bbl
\begin{thebibliography}{10}

\bibitem{tarokh}
V.~Tarokh, H.~Jafarkhani, and A.~R. Calderbank, ``Space-time block codes from
  orthogonal designs,'' {\em IEEE Trans. Inform. Theory}, vol.~45,
  pp.~1456--1466, Jul 1999.

\bibitem{gamal}
H.~E. Gamal and M.~O. Damen, ``Universal space-time coding,'' {\em IEEE Trans.
  Inform. Theory}, vol.~49, pp.~1097--1119, May 2003.

\bibitem{sethuraman}
B.~A. Sethuraman, B.~S. Rajan, and V.~Shashidhar, ``Full-diversity, high-rate
  space-time block codes from division algebras,'' {\em IEEE Trans. Inform.
  Theory}, vol.~49, pp.~2596--2616, Oct 2003.

\bibitem{heath}
R.~W. Heath and A.~J. Paulraj, ``Linear dispersion codes for {MIMO} systems
  based on frame theory,'' {\em IEEE Trans. on Signal Processing}, vol.~50,
  pp.~2429--2441, Oct 2002.

\bibitem{wolniansky}
P.~W. Wolniansky, G.~J. Foschini, G.~D. Golden, and R.~A. Valenzuela,
  ``{V-BLAST}: An architecture for realizing very high data rates over the
  rich-scattering wireless channel,'' {\em Proc. Int. Symp. Signals, Systems,
  and Electronics}, pp.~295--300, Oct 1998.

\bibitem{barbarossa}
S.~Barbarossa, {\em Multiantenna Wireless Communication Systems}.
\newblock Artech House, Norwood, 2005.

\bibitem{sendonaris}
A.~Sendonaris, E.~Erkip, and B.~Aazhang, ``User cooperation diversity - part
  {I} and part {II},'' {\em IEEE Trans. Commun.}, vol.~51, pp.~1927--48, Nov
  2003.

\bibitem{laneman}
J.~Laneman and G.~Wornell, ``Distributed space-time-coded protocols for
  exploiting cooperative diversity in wireless networks,'' {\em IEEE Trans. on
  Inform. Theory}, vol.~49, pp.~2415--2425, Oct 2003.

\bibitem{jing}
Y.~Jing and B.~Hassibi, ``Distributed space-time coding in wireless relay
  networks,'' {\em IEEE Trans. on Inform. Theory}, vol.~49, pp.~3524--3536, Dec
  2006.

\bibitem{borade}
S.~Borade, L.~Zheng, and R.~Gallager, ``Amplify-and-forward in wireless relay
  networks: Rate, diversity, and network size,'' {\em IEEE Trans. on Inform.
  Theory}, vol.~53, pp.~3302--3318, Oct 2007.

\bibitem{hunter}
T.~E. Hunter and A.~Nosratinia, ``Diversity through coded cooperation,'' {\em
  IEEE Trans. Wireless Commun.}, vol.~5, pp.~283--289, Feb 2006.

\bibitem{shalom}
Y.~Bar-Shalom and X.-R. Li, {\em Estimation and Tracking}.
\newblock Artech House, London, 1993.

\bibitem{nielsen}
R.~O. Nielsen, {\em Sonar Signal Processing}.
\newblock Artech House, London, 1991.

\end{thebibliography}
